\documentclass[12pt]{iopart}

\usepackage{graphicx}
\usepackage{geometry}
\usepackage{listings}
\usepackage{xcolor}
\usepackage{array}
\usepackage{caption}
\usepackage{subcaption}
\usepackage{hyperref}

\usepackage{arydshln}
\usepackage{makecell}
\usepackage{pdflscape}

\newcommand{\gw}{gravitational wave}
\newcommand{\gws}{gravitational waves}

\newcommand{\scl}{scattered-light}
\newcommand{\Scl}{Scattered-light}

\begin{document}

\title[ArchEnemy]{ArchEnemy: Removing \scl{} glitches from gravitational wave data.}

\author{Arthur E. Tolley$^{1}$, Gareth S. Cabourn~Davies$^{2}$, Ian W. Harry$^{2}$ and Andrew P. Lundgren$^{2}$}

\address{$^{1}$DISCnet Centre for Doctoral Training,
         Institute of Cosmology and Gravitation,
         University of Portsmouth,
         Portsmouth PO1 3FX,
         United Kingdom}
\address{$^{2}$
         Institute of Cosmology and Gravitation,
         University of Portsmouth,
         Portsmouth PO1 3FX,
         United Kingdom}

\ead{arthur.tolley@port.ac.uk}  

\vspace{10pt}
\begin{indented}
\item[]\today
\end{indented}

\begin{abstract}
Data recorded by \gw{} detectors includes many non-astrophysical transient noise bursts, the most common of which is caused by \scl{} within the detectors. These so-called ``glitches'' in the data impact the ability to both observe and characterize incoming \gw{} signals. In this work we use a \scl{} glitch waveform model to identify and characterize \scl{} glitches in a representative stretch of \gw{} data. We identify $2749$ \scl{} glitches in $5.96$ days of LIGO-Hanford data and $1306$ glitches in $5.93$ days of LIGO-Livingston data taken from the third LIGO-Virgo observing run. By subtracting identified \scl{} glitches we demonstrate an increase in the sensitive volume of a \gw{} search for binary black hole signals by $\sim1\%$.
\end{abstract}




\newpage
\section{\label{sec:intro}Introduction}

The Laser Interferometer Gravitational-Wave Observatory (LIGO)~\cite{aligo} and Virgo~\cite{avirgo} collaborations made the first observation of \gws{} in September 2015~\cite{first_detection}. The detection established the field of \gw{} astronomy and a global network of \gw{} detectors, now joined by KAGRA~\cite{kagra}, has allowed for the detection of approximately 100 \gw{} events~\cite{gwtc1, gwtc2, gwtc21, gwtc3}.

The detection of \gws{} is made possible by both the sensitivity of the detectors and the search pipelines~\cite{pycbc, gstlal, spiir, mbta, cwb_2020} which analyse raw strain data from the output of the detectors and identify observed \gw{} signals. One of the problems that these search pipelines must deal with is the fact the data contains both non-stationary noise and short duration `glitches'~\cite{GuideToDetNoise,DetCharO2O3, VirgoDetChar} where noise power increases rapidly. Glitches are caused by instrument behaviour or interactions between the instrument and the environment~\cite{transient_noise, Glanzer:2023hzf} and glitches reduce the sensitivity of the detectors~\cite{sensitivity_o3}, can potentially obscure candidate \gw{} events~\cite{gwtc2} and can even mimic \gw{} events~\cite{GWMimicking, PyCBC_Singles}.

\begin{figure}
  \includegraphics[width=\textwidth]{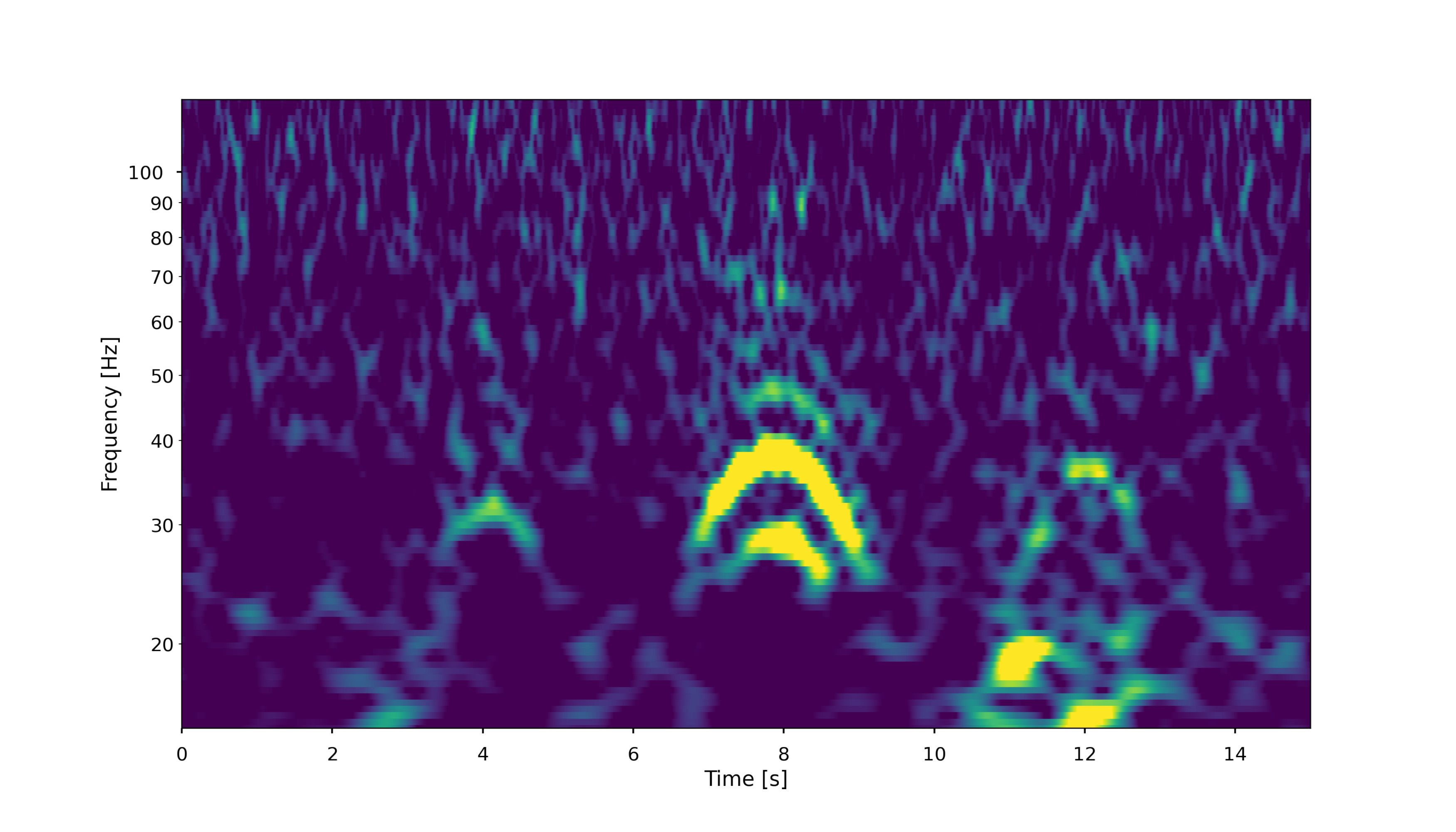}
  \caption{An Omega scan \cite{omegascan} of \gw{} data containing an example of a \scl{} glitch. \Scl{} glitches are characterized by a symmetric arch-like pattern. Multiple \scl{} glitches can be seen at the 4, 8 \& 12 second marks as well as multiple harmonic glitches at 8 seconds.}
  \label{fig:scattered_light}
\end{figure}

Different classes of glitches have been characterized using tools such as Gravity Spy~\cite{gravityspy, GSpy_update}. Of the 325,101 glitches classified by Gravity Spy in the third observing run of Advanced LIGO~\cite{GSpy2022} with a confidence of $90\%$ or higher, 120,733 ($32.1\%$) were classified as ``Scattered Light''. \Scl{} glitches occur in the 10-120Hz frequency band~\cite{reducing_scattering_o3} which coincides with the frequency band where we observe the inspiral and merger signatures of compact binary coalescences. \Scl{} glitches are characterized by an arch-like pattern in a time-frequency spectrogram of the detector output, as seen in figure~\ref{fig:scattered_light}. 
\Scl{} glitches occur when laser light in the interferometer is scattered from the main optical path by components within the detector. The motion of these components is coupled to seismic motion inducing a phase shift on the light being scattered as the surface moves back and forth. This \scl{} then recombines with the main laser, producing \scl{} glitches in the data. The surfaces from which \scl{} glitches originate have been objects on optical benches such as lenses, mirrors and photo-detectors~\cite{TAccadia}.

\begin{figure}
  \includegraphics[width=\textwidth]{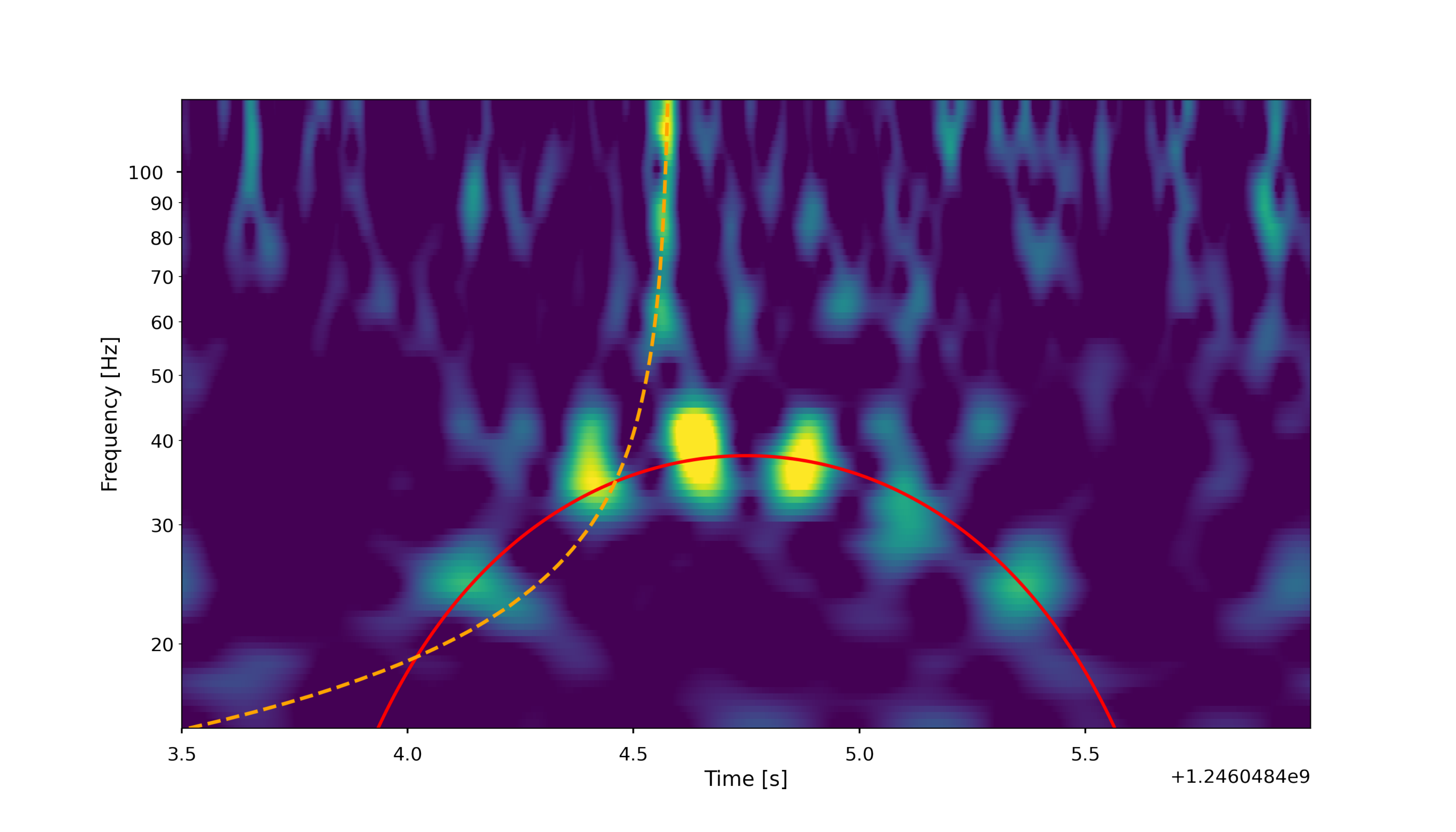}
  \caption{GW190701\_203306, a \gw{} event coincident with a \scl{} glitch in the data from the LIGO Livingston observatory. The orange dashed track shows the inferred time-frequency evolution of a \gw{} event produced by a compact binary merger, the red solid line is an overlaid track of the approximate location of the coincident fast scattering glitches.}
  \label{fig:obscured_detection}
\end{figure}

\begin{figure}
  \includegraphics[width=\textwidth]{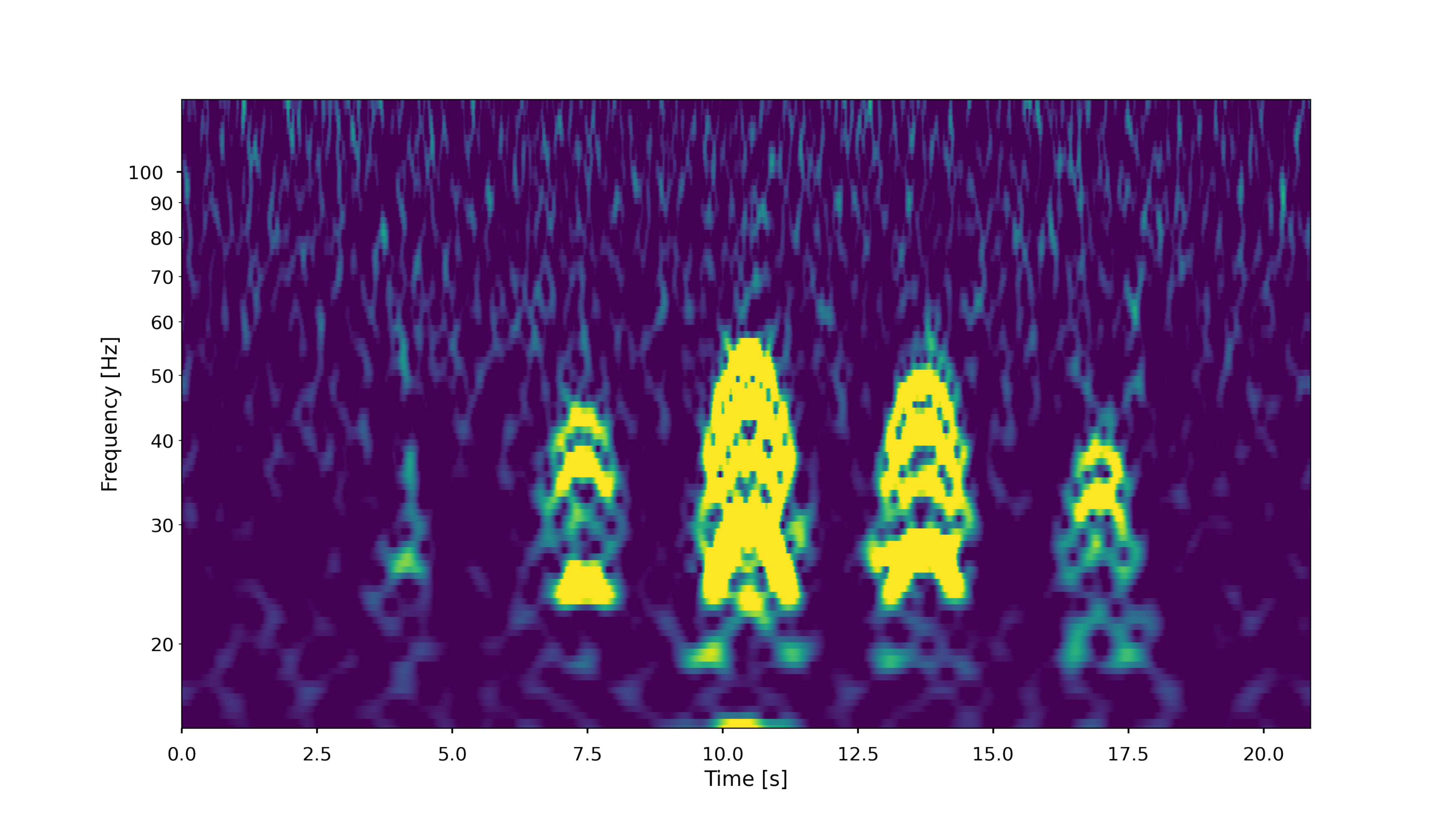}
  \caption{An Omega scan \cite{omegascan} of \gw{} data containing multiple examples of \scl{} glitches. Here we see multiple \scl{} glitches repeating periodically in a $20$ second period of time. Harmonic \scl{} glitches are also seen in multiple stacks, the harmonic glitches share \emph{time period} values and the \emph{glitch frequency} values are $n-$multiples of the lowest frequency harmonic within the stack.}
  \label{fig:consec_scattered_light}
\end{figure}

\Scl{} glitches have been a significant problem when observing compact binary mergers. As an example, GW190701\_203306 was coincident with a \scl{} glitch, as shown in figure \ref{fig:obscured_detection}~\cite{gwtc2}, requiring subtraction from the data before the event could be properly characterized~\cite{O3_subtraction}. A further 7 candidate events were found to be in coincidence with \scl{} glitches in the third observing run~\cite{gwtc3}. For this reason, it is important to reduce the effect of \scl{} glitches in the detectors and \gw{} search pipelines. \Scl{} glitches occur as single or multiple glitches and can appear rapidly in time and simultaneously in frequency (see figure \ref{fig:consec_scattered_light}), which we refer to as harmonic glitches.

The most obvious way to remove the impact of glitches is removing the mechanisms which produce the glitches in the observatories. This has been investigated in previous works~\cite{reducing_scattering_o3, TAccadia, Laura_Noise, gwadaptive, HilbertHuang, tvf-EMD, Longo_daily, MichalSub}, which focus on identifying the surfaces in which light is being scattered from and then mitigating the scattering by reducing the reflectivity of the surface, seismically isolating it or relocating it. 

An alternative method for reducing \scl{} glitches, known as ``RC tracking'', was implemented in the Advanced LIGO observatories in January 2020~\cite{reducing_scattering_o3}. The Advanced LIGO detectors employ a quadruple pendulum suspension for the test masses where two chains suspend four masses in this suspension system, one for the test mass optic and the other for the reaction mass. The reaction mass is used to impose a force upon the test mass and a significant source of \scl{} glitches was the large relative motion between the test mass chain and the reaction chain. To mitigate this effect, the relative motion between the end test mass and the reaction mass needed to be reduced. This was achieved by ensuring the two chains are moving together by applying force to the top stage of the quadruple suspension system in Advanced LIGO. The implementation of RC tracking represented a decrease from $0.01 s^{-1}$ to $0.0001 s^{-1}$ and $0.0072 s^{-1}$ to $0.0012 s^{-1}$ in the number of \scl{} glitches detected by Gravity Spy for LIGO-Hanford and LIGO-Livingston respectively.

While methods for preventing \scl{} glitches have been developed and have shown success, they have not been able to remove the problem of \scl{} glitches from the data. Additionally, as the detectors continue to be upgraded and increase in sensitivity, new sources of \scl{} glitches will continue to appear. Identifying these new sources and mitigating their effects can take many months, during which time the detectors are taking in data which might be affected by the presence of \scl{} glitches. Therefore, it is not realistic to believe that analyses will be able to regularly run on data that does not contain \scl{} glitches and this motivates us to develop a technique for mitigating the impact of these glitches when trying to identify compact binary mergers in \gw{} data.

In this work we present a new method for identifying and removing \scl{} glitches from \gw{} data in advance of running searches to identify compact binary mergers.
We first introduce a method for identifying when \scl{} glitches are present in detector data, through the creation of a new modelled search for \scl{} glitches, similar to how we search for \gws{} using matched filtering. We can model \scl{} glitches, generate a suitable set of glitch waveforms and perform a matched filter search on detector data. We then subtract identified glitches from the data to increase detector sensitivity. The detector data isn't Gaussian and stationary so the matched filter does have the potential to identify non-\scl{} glitches, and potentially even \gw{} signals, as \scl{} glitches. To prevent this we also demonstrate a new \scl{} $\chi^{2}$ test, which can distinguish between \scl{} glitches, and other glitches---and \gw{} signals---in the data.

We begin by reviewing previous research and describing the formulation of the waveform model used for characterizing \scl{} glitches in section \ref{sec:sc_li}. In section \ref{sec:search_techniques} we introduce the various techniques used in the search to identify \scl{} glitches in \gw{} data and the results of the \scl{} glitch search. In section \ref{sec:results} we describe the results of a ``glitch-subtracted'' \gw{} search and any increases in sensitivity. We conclude in section \ref{sec:conclusion} and discuss the implementation of this method in future observing runs.

\section{\label{sec:sc_li}\Scl{}}

To identify \scl{} glitches in \gw{} data requires an accurate model of \scl{} glitches. This section details the derivation of the model we will use for generating our \scl{} glitch filter waveforms, along with its parameterization.

\subsection{Modelling \scl{} glitches - a review}

Our model for \scl{} glitches draws heavily from~\cite{TAccadia}, and we briefly review the main details of the model presented there. In~\cite{TAccadia}, the authors construct a model to accurately predict the increase in noise due to \scl{} during periods of increased micro-seismic activity. The model in~\cite{TAccadia} is constructed from parameters related to physically measurable properties of the detector such as the mirror transmission factor, $T$, the finesse of the Fabry–Pérot cavity, $F$, or the wavelength of the light, $\lambda$.

They define the amplitude of the additional beam produced by light scattering off of a surface as
\begin{equation}
    A_{sc} = A_{0} T \sqrt{\frac{2 F}{\pi}} \sqrt{f_{sc}} e^{i \phi_{sc}}\,,
    \label{eqn:accadia_amplitude}
\end{equation}

where $A_{0}$ is the amplitude of the light resonating in a Fabry–Pérot cavity, $f_{sc}$ is the fraction of the optical power scattered back into the main beam and $\phi_{sc}$ is the phase angle modulated by the displacement of the scattering optics, defined as
\begin{equation}
    \phi_{sc}(t) = \frac{4 \pi}{\lambda} ( x_{0} + \delta x_{opt}(t) ),
    \label{eqn:accadia_phase_noise}
\end{equation}
where $\delta x_{opt}$ is the displacement of the scattering surface and $x_0$ is the static optical path.

The total amplitude of the beam inside the arm is given by $A_{tot} = A_{0} + A_{sc}$, with a phase angle equal to the phase noise introduced by the \scl{} $\delta \Phi = \frac{A_{sc}}{A_{0}} \cdot \sin \phi_{sc}$. The noise introduced by the \scl{}, $h_{sc}$, can be approximated through the relationship $\sin(\Phi_0+ \delta\Phi) \approx \cos(\Phi_0) \times \sin(\delta \Phi)$ for small $\delta\Phi$, and can be expressed as
\begin{equation}
    h_{sc}(t) = G \cdot \sin \left(\frac{4 \pi}{\lambda} (x_{0} + \delta x_{sc}(t) ) \right),
    \label{eqn:accadia_strain}
\end{equation}
where $G$ is the \emph{coupling factor}, defined as $K \cdot \sqrt{f_{sc}}$ where
\begin{equation}
K = \frac{\lambda}{4 \pi} \frac{T}{\sqrt{2 F \pi}}.
\end{equation}

The displacement of the scatterer when presenting with oscillatory motion is then given as
\begin{equation}
    \delta x_{sc} (t) \simeq A_{m} \sin(2 \pi f_{m} t),
    \label{eqn:accadia_oscillatory}
\end{equation}
where $f_{m}$ is the frequency-modulated signal with modulation index $m = A_{m} \frac{4 \pi}{\lambda}$ and where $A_{m}$ is the amplitude of the $n$th harmonic. Finally, equation~\ref{eqn:accadia_strain} can be simplified when considering only small bench motion, according to
\begin{equation}
    h_{sc}(t) = G \cdot \cos\phi_{0} \cdot \frac{4 \pi}{\lambda} \cdot \delta x_{sc}(t) \;.
    \label{eqn:accadia_strain_linearized}
\end{equation}

\subsection{Model}

The model introduced in~\cite{TAccadia} for the \gw{} strain noise introduced by \scl{} uses a lot of knowledge about the detector state. The model used in this work will be more phenomenological in the parameterization, allowing us to rely only on the characteristics of the glitches in the strain data and not knowledge of the detector configuration, especially in cases where this detector information might not be known. Each \scl{} glitch, as viewed in a spectrogram (see figure \ref{fig:scattered_light}), has two easily identifiable features: the maximum frequency reached, \emph{glitch frequency} ($f_{gl}$); and the period of time the glitch exists in detector data, \emph{time period} ($T$).  In addition we can fully describe an artifact by defining an \emph{amplitude} ($A$), \emph{phase} ($\psi$) and \emph{center time} of the glitch ($t_0$).

To formulate a model of \scl{} glitches in terms of these parameters, we simplify equation \ref{eqn:accadia_strain_linearized}, treating the strain noise caused by \scl{} as the sinusoidal function
\begin{equation}
  h_{sc}(t) \propto \sin(\phi_{noise}(t)).
  \label{eqn:h_sc_initial}
\end{equation}
Here the induced phase noise ($\phi_{noise}$) is equal to 
\begin{equation}
    \phi_{noise}(t) = 2 \pi f_{rep} t,
     \label{eqn:phi_noise}
\end{equation}
and $f_{rep}$ is the frequency of repetition of the sinusoid and is directly related to the \emph{time period}, $T$,
\begin{equation}
  f_{rep} = \frac{1}{2 T}.
  \label{eqn:f_rep}
\end{equation}
The \emph{time period} of a \scl{} glitch only corresponds to half of a sinusoidal wave hence the multiplier of 2 on the denominator of equation \ref{eqn:f_rep}.

\Scl{} glitches are caused by the physical increase in the distance travelled by the light as a consequence of being reflected off of a surface. The light returning to the beamsplitter from one arm will have travelled a different path length compared to the other arm and this path difference will act as a phase difference between the two arms causing non-destructive interference. The path difference and phase difference can be related with
\begin{equation}
  \Delta \phi_{scattering}(t) = 2 \cdot \frac{2 \pi}{\lambda} \Delta x(t),
  \label{eqn:delta_phi_scat}
\end{equation}
where $\Delta x(t)$ indicates the change in the path taken by the light over time. An additional multiplier of 2 indicates our path difference is occurring twice, once when the light approaches the surface and again when leaving.

We assume the scattering surfaces are oscillatory in motion and apply the same simplification made in equation \ref{eqn:accadia_oscillatory}, with an initial position $\Delta x = 0$ to a maximum displacement of $x_{0}$. This produces an equation for the path difference of the light,
\begin{equation}
  \Delta x(t) = x_{0} \sin(2 \pi f_{rep} t).
  \label{eqn:delta_x_t}
\end{equation}
We substitute equation \ref{eqn:delta_x_t} into equation \ref{eqn:delta_phi_scat} and produce an equation for the phase noise induced by \scl{}
\begin{equation}
  \phi_{scattering}(t) = \frac{4 \pi}{\lambda} x_{0} \sin(2 \pi f_{rep} t) \;,
  \label{eqn:phi_sc}
\end{equation}
this equation for the phase noise induced by \scl{}, $\phi_{scattering}(t)$, and the equations for the generic phase noise, $\phi_{noise}(t)$ (equation \ref{eqn:phi_noise}), can now be used to create an equation for the strain noise caused by \scl{}.

We take the derivatives of equations \ref{eqn:phi_noise} $\&$ \ref{eqn:phi_sc} with respect to time:
\begin{equation}
  \phi_{noise} ^\prime (t) = 2 \pi F_{inst} (t),
\end{equation}
\begin{equation}
  \phi_{scattering} ^\prime (t) = \frac{4 \pi}{\lambda} x_{0} 2 \pi f_{rep} \cos(2 \pi f_{rep} t) \;,
\end{equation}
where $F_{inst}$ is the instantaneous frequency at a specific time. We generate \scl{} glitches from $\frac{-T}{2}$ to $\frac{T}{2}$ to ensure their maximum frequency occurs at $t = 0$. We define this maximum frequency as the \emph{glitch frequency}.

We equate the two derivatives and set $t = 0$, this replaces $F_{inst}(t)$ with $f_{gl}$ and $\cos(2\pi f_{rep} t)$ with $1$, then re-arrange to find the maximum displacement of the scattering surface, $x_{0}$, as
\begin{equation}
  x_{0} = \frac{f_{gl}}{f_{rep}} \frac{\lambda}{4 \pi}.
  \label{eqn:x0}
\end{equation}
Substituting equation \ref{eqn:x0} into equation \ref{eqn:phi_sc} gives us an expression for the \scl{} phase noise,
\begin{equation}
  \phi_{scattering}(t) = \frac{f_{gl}}{f_{rep}} \sin(2 \pi f_{rep} t),
  \label{eqn:phi_sc2}
\end{equation}
and substituting equation \ref{eqn:phi_sc2} as our $\phi_{noise}$ in equation \ref{eqn:h_sc_initial} we arrive at our model of \scl{} glitches,
\begin{equation}
  h_{sc}(t) = A \sin\left(\frac{f_{gl}}{f_{rep}} \sin(2 \pi f_{rep} t) + \psi\right),
  \label{eqn:model}
\end{equation}
where $A$ and $\psi$ are amplitude and phase parameters to be maximised over.

In~\cite{MichalSub} they provide another term to describe \scl{} glitches which uses the instantaneous frequency as a function of time, allowing the identification of the correct amplitude at all frequencies. This term is due to radiation pressure coupling and is thought to be dominant at low frequencies. The relationship between our amplitude and the new amplitude depends on the power in the arm cavities and the signal recycling mirror reflectivity which we do not consider in our model and so disregard this extra term.

\subsection{Harmonics}

As described in~\cite{TAccadia}, harmonic glitches appear at the same time with different \emph{glitch frequencies}. A harmonic glitch is a glitch with a \emph{glitch frequency} that is a positive integer multiple of the glitch frequency of the glitch in the stack with the lowest glitch frequency value. This lowest frequency glitch has the potential to appear below $15$Hz and will be masked by other sources of noise and therefore cannot be seen. An example of harmonics can be seen in figure \ref{fig:consec_scattered_light}.

\section{\label{sec:search_techniques}Identifying \scl{} glitches in \gw{} strain data}

Equipped with our model for \scl{} glitches from the previous section, we now discuss how we identify and parameterize \scl{} glitches in a stretch of \gw{} data before we apply this to searches for compact binary mergers in the next section. 

\subsection{Matched Filtering \label{subsec:MF}}

Given our model of \scl{} glitches, we can consider a realization with specific values of the parameters discussed above. To identify glitches with these values of the parameterization, we follow the same technique as that commonly used to identify compact binary mergers and apply matched filtering. The matched filter is defined as~\cite{findchirp}. 
\begin{equation}
  \rho(t) = \frac{(h | s)}{\sqrt{(h | h)}} \equiv (\hat{h} | s).
  \label{eqn:mf_1}
\end{equation}
Here $h$ is the model template, $s$ the \gw{} data we are searching and we use the noise-weighted inner product defined between two time series $a(t)$ and $b(t)$ as
\begin{equation}
  (a | b) = 4 Re \int^{\infty}_{0} \frac{\tilde{a}(f) \tilde{b}^*(f)}{S_n(f)} 
  df.
  \label{eqn:inner_product}
\end{equation}
The tilde on $\tilde{a}$ and $\tilde{b}$ refer to the Fourier-transform of both variables into the frequency domain. The denominator, $S_n(f)$, represents the one-sided power spectral density (PSD) of the data, defined as
\begin{equation}
  \langle \tilde{s}(f) \tilde{s}(f^\prime) \rangle = \frac{1}{2} S_n(f) \delta(f - f^\prime) \;,
  \label{eqn:psd}
\end{equation}
where the angle brackets denote an average over noise realizations and $\delta$ is the Dirac delta function.

\Scl{} glitches will take a range of values of the parameters describing them and we must be able to identify glitches anywhere in the parameter space. Following~\cite{findchirp} we can analytically maximize over the \emph{amplitude}, \emph{phase} and the \emph{center time} glitch parameters in equation \ref{eqn:model}. The matched filter naturally maximizes over \emph{amplitude} when expressed in terms of signal-to-noise ratio, and can be evaluated as a function of time by including a time shift in equation~\ref{eqn:inner_product},
\begin{equation}
  (a | b)(t) = \left| 4 Re \int^{\infty}_{0} \frac{\tilde{a}(f) \tilde{b}^*(f)}{S_n(f)} 
  e^{-2 \pi i f t} 
  df \right|.
  \label{eqn:inner_product_time}
\end{equation}
To maximize over phase we take the absolute value of the inner product, rather than the real part of the integral.

\subsection{Template Bank}

Our \scl{} glitch model is parameterized by 5 variables. As shown above, by maximizing over \emph{phase}, \emph{amplitude} and \emph{time}, we can analytically maximize the signal-to-noise ratio over 3 of these variables. However, the remaining two describe the intrinsic evolution of the glitch and we must explicitly search over these parameters. To do this we create and use a ``template bank'' of \scl{} glitch waveforms, created such that it would be able to identify glitches with any value of our 2 remaining parameters, \emph{glitch frequency} and \emph{time period}.

To generate this template bank, a stochastic template placement algorithm~\cite{template_bank_placement} is used. This algorithm randomly generates a new template, places the template in the existing template bank and calculates the ``distance'' (how similar two templates appear) between the new template and existing templates. If the new template is too close to an existing template it is discarded, otherwise, it is accepted into the bank. The density of the bank is dependent on the allowed distance between two templates. The dominant cost of the search for \scl{} glitches is matched filtering and is approximately linear in the number of templates, a larger template bank will find all of the glitches with more accurate parameter values but the computational cost of the search will be increased. The template bank generation alone does not constitute a significant computational cost in the search.

The distance function we have chosen to evaluate templates by is the match between two glitch templates. To calculate the match, we first normalize both templates such that the matched filter between a template and itself would be equal to 1. We can then compute the inner product between the two normalized templates to find their match
\begin{equation}
    M = \max_{t, \psi} (\hat{a} | \hat{b}).
  \label{eqn:match}
\end{equation}
The value for the match, $M$,  is bounded between 0 and 1, where a value of 0 indicates orthogonal waveforms and a value of 1 indicates identical normalized waveforms. The maximum match allowed between any two templates in our bank is 0.97, which implies that a fully converged stochastic bank would have at least one waveform in it with $M \geq 0.97$ for any point in the space of parameters. 

\begin{figure}
  \centering
  \includegraphics[width=0.7\textwidth]{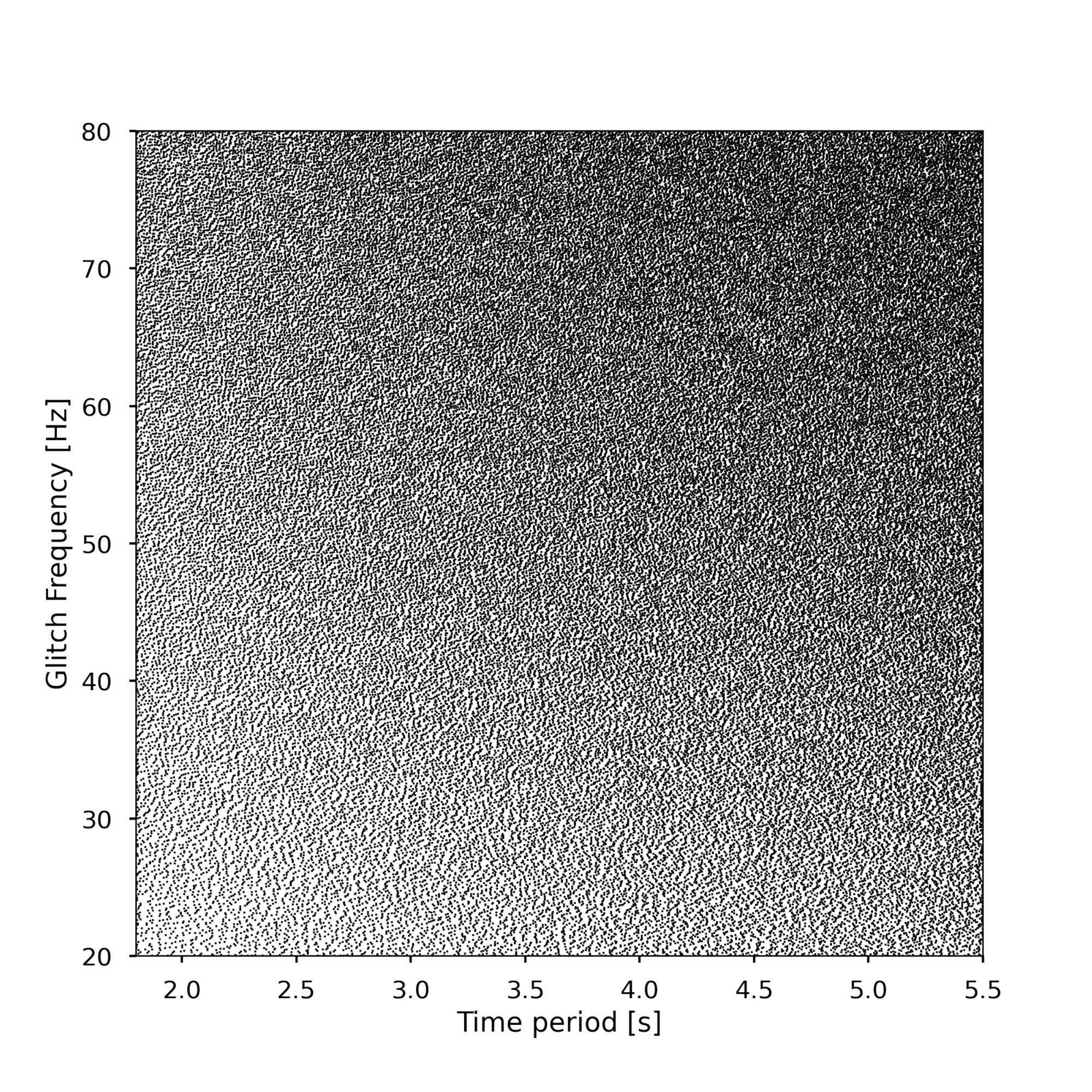}
  \caption{The template bank of \scl{} glitch templates (black points) used in the search for \scl{} glitches. The \emph{glitch frequency} parameter values range from $20$Hz - $80$Hz and the \emph{time period} parameter values range from $1.8$s - $5.5$s. This bank was created with a maximum match allowed between two templates of $0.97$ and contains 117,481 templates.}
  \label{fig:sq_bank}
\end{figure}

For our \scl{} glitch search, we generated a template bank with \emph{time periods} $\in$ $1.8$s - $5.5$s and \emph{glitch frequencies} $\in$ $20$Hz - $80$Hz. We chose to use the Advanced LIGO zero-detuned high-power sensitivity curve~\cite{SensCurve}\footnote{The zero-detuned high-power sensitivity curve is a broader noise curve than the O3 Advanced LIGO data that we identify \scl{} glitches in. However, this broader noise curve will result in \emph{more} template waveforms that we need, and will therefore overcover the parameter space, rather than potentially miss \scl{} glitches.}. This allowed us to generate a template bank that contains 117,481 templates. We visualize the distribution of the templates as a function of \emph{time period} and \emph{glitch frequency} in figure~\ref{fig:sq_bank}, observing a greater density of templates at higher frequencies and longer durations.

\subsection{Identifying potential \scl{} glitches in the data}

We test our method by searching through \gw{} data from 2019-11-18 16:13:05 - 2019-11-25 22:11:09 for the LIGO-Hanford and LIGO-Livingston detectors. This time corresponds to the 25th period of data used in the LVK analysis of O3 data for compact binary mergers~\cite{gwtc3} and is prior to the implementation of RC tracking~\cite{reducing_scattering_o3} in these detectors. We only analyse data that is flagged as ``suitable for analysis" on the Gravitational Wave Open Science Center~\cite{GWOSC}\footnote{We note that in O3 \emph{only} data suitable for analysis is released, so we simply analyse all of the publicly available data.}. This corresponds to 5.96 days of analysable data for LIGO-Hanford and 5.93 days for LIGO-Livingston.

Equipped with our template bank we now identify potential \scl{} glitches in the data. We matched filter all of the data against all of the templates, producing a signal-to-noise ratio time series for every template in the bank. These signal-to-noise ratio time series will contain peaks which, when above a certain limit, indicate the presence of a \scl{} glitch at a particular time. We retain any maxima within the time series that have a signal-to-noise ratio greater than 8. However, as we do this independently for every template, we will identify multiple peaks for any given glitch, and we will also find peaks that correspond to other types of glitch, or even \gw{} signals. We discuss how we reduce this to a list of identified \scl{} glitches in the following subsections.

\subsection{Scattered-Light Signal Consistency Test}

To prevent the search for \scl{} glitches from misclassifying other classes of glitches, or \gw{} signals, we employ a $\chi^2$ consistency test. The $\chi^2$ discriminator introduced in \cite{Allen_2005} divides \gw{} templates into a number of independent frequency bins. These bins are constructed so as to contain an equal amount of the total signal-to-noise ratio (SNR) of the original matched filter between template and data. The $\chi^{2}$ value is obtained by calculating the SNR of each bin, subtracting this from the expected SNR value for each bin and squaring the output. These values are summed for all bins and this value forms the $\chi^{2}$ statistic,
\begin{equation}
  \chi_{r}^{2} = \frac{\chi^{2}}{\textrm{DOF}} = \frac{n}{2n - 2} \sum_{i=1}^n \left(\frac{\rho}{\sqrt{n}} - \rho_{bin,i}\right)^2.
  \label{eqn:chi_squared}
\end{equation}
Here $n$ is the number of bins, $\rho$ is the SNR of the original matched filter between template and data, and $\rho_{bin}$ is the value of the SNR found when matched filtering one of the divided templates and the data. This test is constructed so as to produce large values when the data contains a glitch, or astrophysical signal, that is not well described by the template, but to follow a $\chi^2$ distribution if a glitch that matches well to the template is present, or if the data is Gaussian and stationary.

\begin{figure}
  \centering
  \begin{minipage}[t]{1.0\linewidth}
  \includegraphics[width=0.49\textwidth]{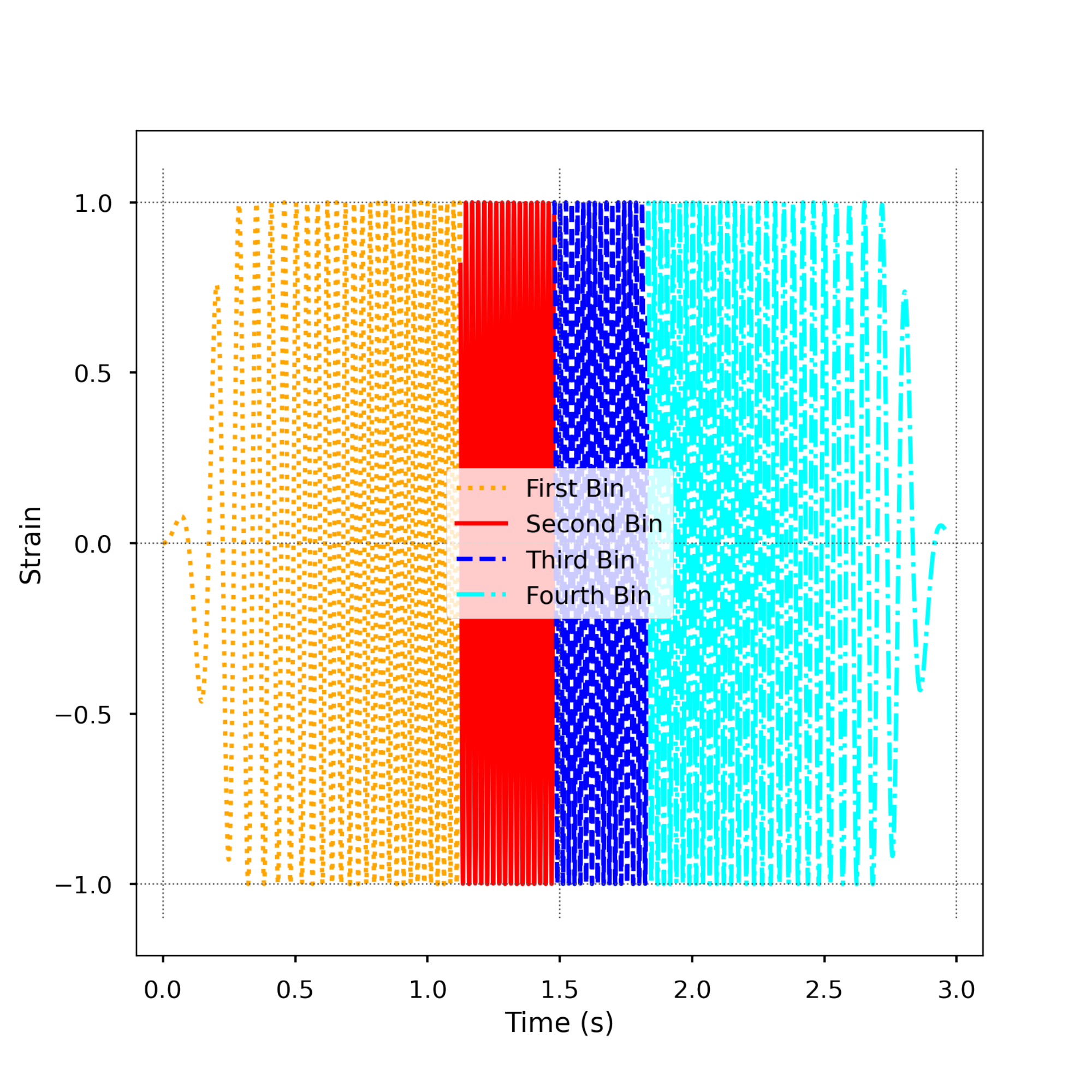}
  \hspace{0.01\linewidth}
  \includegraphics[width=0.49\linewidth]{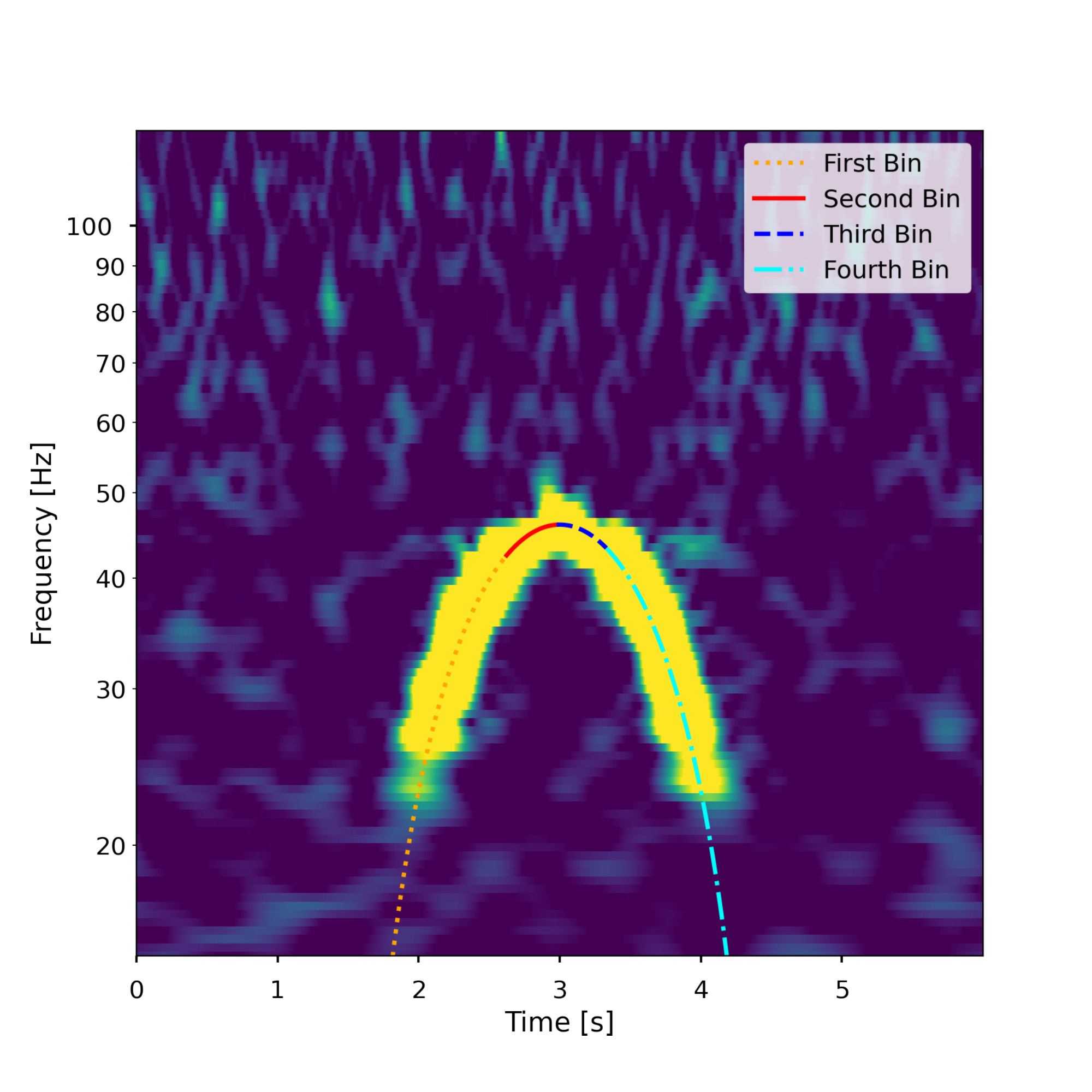}
  \end{minipage}
  \caption{A \scl{} glitch template (left) where the colours and line-styles are indicative of the four equal SNR time bins to be used in calculating the $\chi_{r}^{2}$ value and re-weighting the SNR. The same \scl{} glitch template bins overlaid on an injection of the \scl{} glitch (right). The inner two bins are considerably shorter than the outer two bins which informs us that the center - higher frequency - region of the \scl{} glitch contributes a larger amount to the SNR per unit time than the lower frequency regions.}
  \label{fig:split_temp_subplot}
\end{figure}

The $\chi^2$ test that we employ is similar to that of \cite{Allen_2005}, however compact binary merger waveforms increase in frequency over time whereas \scl{} glitch templates are symmetric about their centre. We therefore choose to construct our $\chi^2$ test with four non-overlapping bins \emph{in the time domain}, each of which contributes equally to the SNR, an example of the split template can be seen in figure \ref{fig:split_temp_subplot}. The matched filter between the bins and data is computed and the $\chi_{r}^{2}$ value is calculated using equation \ref{eqn:chi_squared} (where $n=4$). Any glitch, or astrophysical signal, which does not exhibit symmetric morphology should not fit with this deconstruction of the template, and should result in elevated $\chi_{r}^{2}$ values.

\begin{figure}
  \makebox[\textwidth][c]{\includegraphics[width=\textwidth]{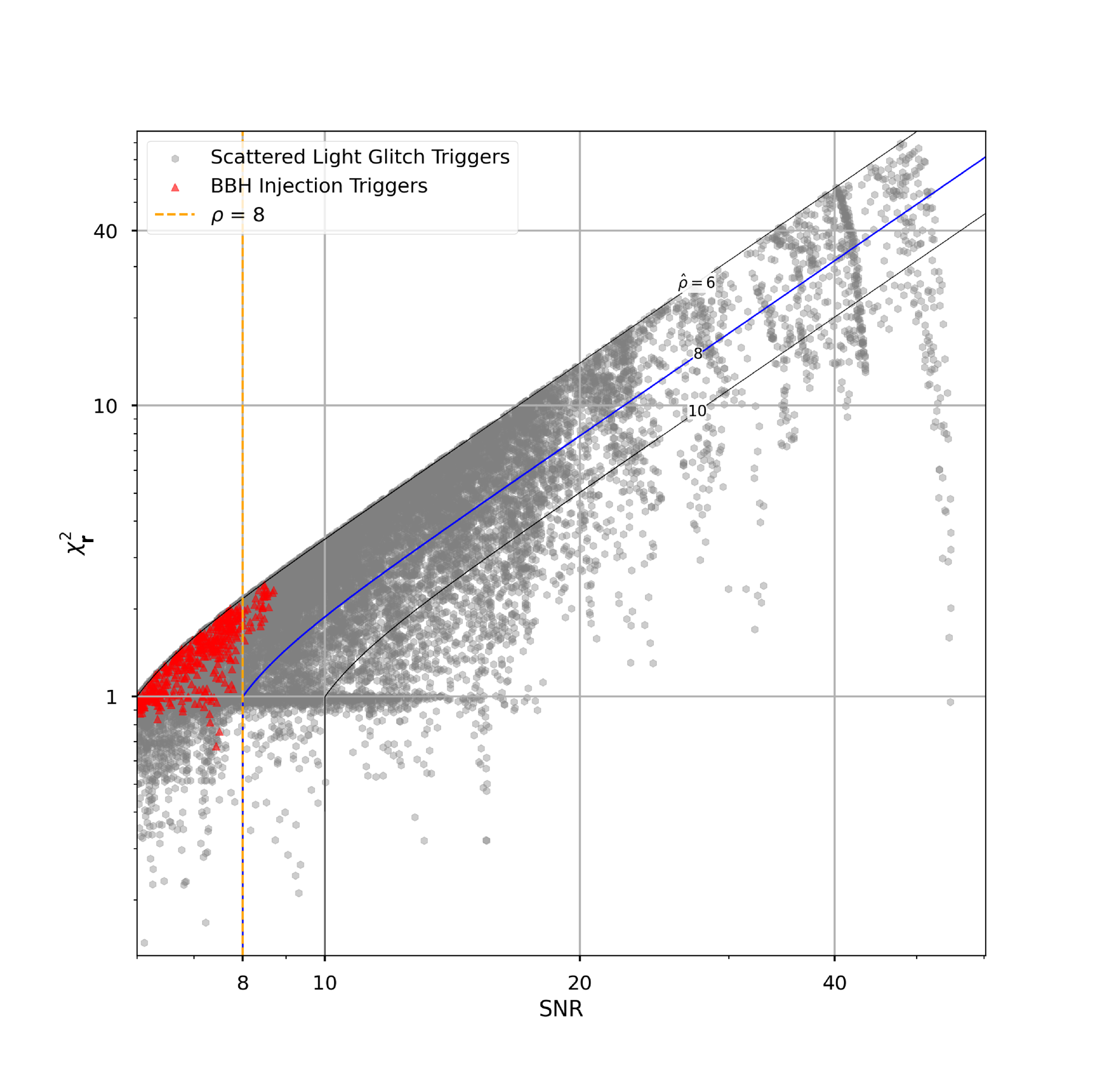}}
  \caption{The signal-to-noise ratio and $\chi_{r}^{2}$ values for the triggers identified by the matched filtering and clustering of the \scl{} glitch template bank with data containing only \scl{} glitches (grey hexagons) and data containing a binary black hole \gw{} injection (red triangles). The black contour lines represent the re-weighted signal-to-noise ratio values the trigger will take when equation \ref{eqn:reweighting} is applied. The orange dashed vertical line indicates the signal-to-noise ratio value limit of 8, above which we decide to perform the $\chi^{2}$ test and calculate the re-weighted signal-to-noise ratio. The blue solid contour line indicates a re-weighted signal-to-noise ratio value of 8, which is the limit at which we consider the trigger to be real. Different re-weighting parameter values will produce different contour line shapes. It can be seen that no triggers for the data containing the \gw{} injection lie beneath the contour line and therefore no \scl{} glitches are found on the \gw{} signal.}
  \label{fig:chi_snr}
\end{figure}

After computing the $\chi_{r}^{2}$ value for potential \scl{} glitches, we follow~\cite{rw_snr_eq} to compute a ``re-weighted signal-to-noise ratio'', which is an empirically tuned statistic depending on the signal-to-noise ratio and the $\chi_{r}^{2}$ value.  The re-weighting function we use matches that presented in~\cite{rw_snr_eq},
\begin{equation}
\rho_{rw} =  \left\{  \begin{array}{l@{\quad}cr} 
\rho & \mathrm{if} & \chi_{r}^{2} \leq 1, \\  
\rho [(1 + (\chi_{r}^{2})^3)/2]^{-\frac{1}{6}} &  \mathrm{if} & \chi_{r}^{2} \ge 1,   
\end{array}\right.
\label{eqn:reweighting}
\end{equation}
where $\rho_{rw}$ represents the re-weighted signal-to-noise ratio of the \scl{} glitch template calculated using the signal-to-noise ratio, $\rho$, and the $\chi_{r}^{2}$ value of that template.
We do note that this re-weighting function has been tuned for compact binary merger waveforms and we did not repeat that tuning here with \scl{} glitches. One could retune this statistic, specifically targeting \scl{} glitches, using (for example) the automatic tuning procedure described in~\cite{McIsaac_2022}. However, we demonstrate the suitability of the $\chi^{2}$ test for our purposes in figure~\ref{fig:chi_snr} where we show the $\chi_{r}^{2}$ vs SNR distribution of the triggers found by our \scl{} glitch search when performed on data containing only \scl{} glitches and data containing a binary black hole \gw{} injection.

We note that the $\chi^{2}$ test increases the number of matched filters required by the search and therefore the computational cost of the search. Each template would require the matched filtering of an additional $4$ ``binned'' templates to calculate the $\chi_{r}^{2}$ value to re-weight the SNR time series of that template, increasing computational cost by a maximum factor of $5$. However, we reduce this increase by only computing the $\chi^{2}$ where it is needed, specifically for any template where the SNR time series has values above the threshold of 8.

\subsection{Identifying all \scl{} glitches in the data}

\begin{figure}
      \begin{minipage}[t]{1.0\linewidth}
        \includegraphics[width=0.49\linewidth]{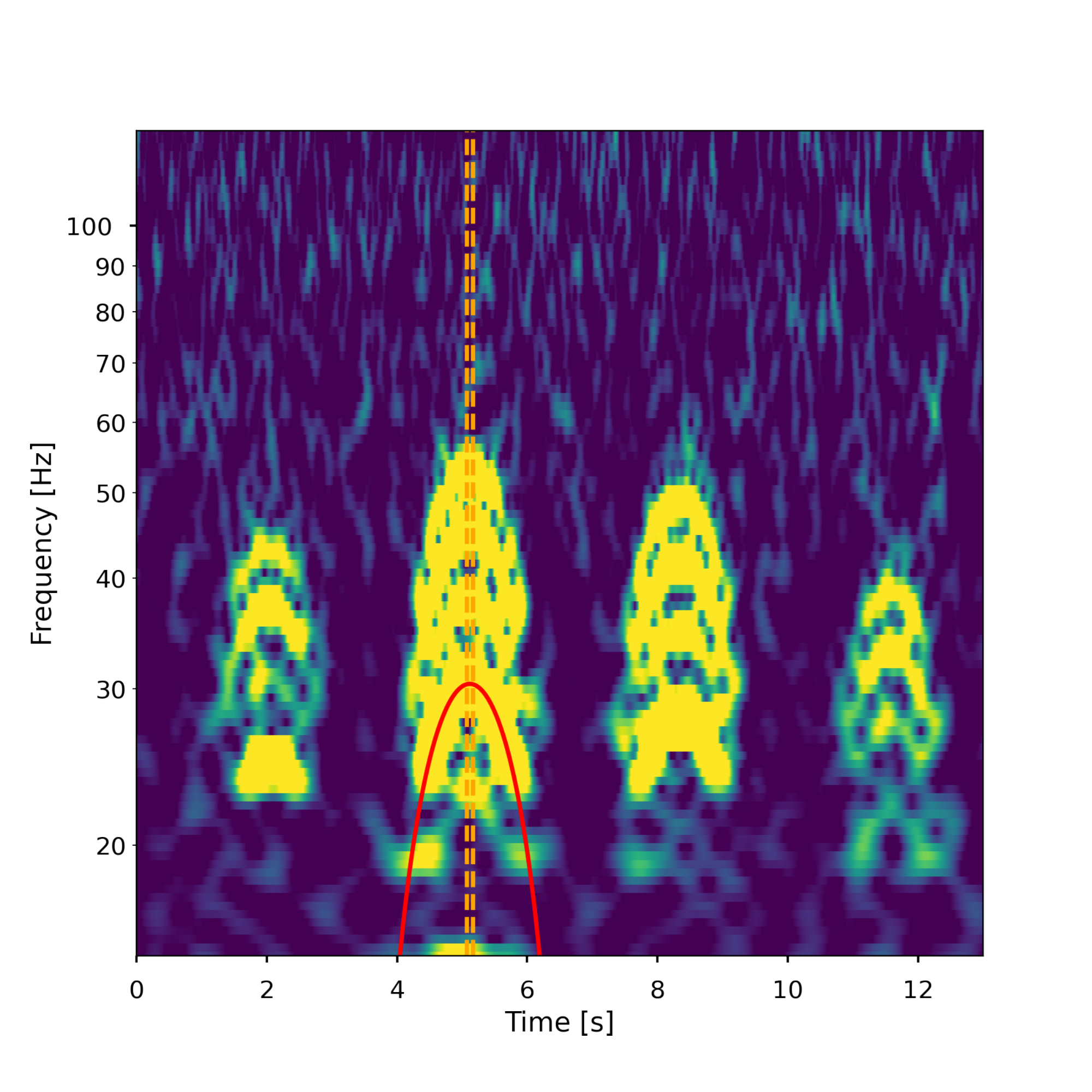}
        \hspace{0.01\linewidth}
        \includegraphics[width=0.49\linewidth]{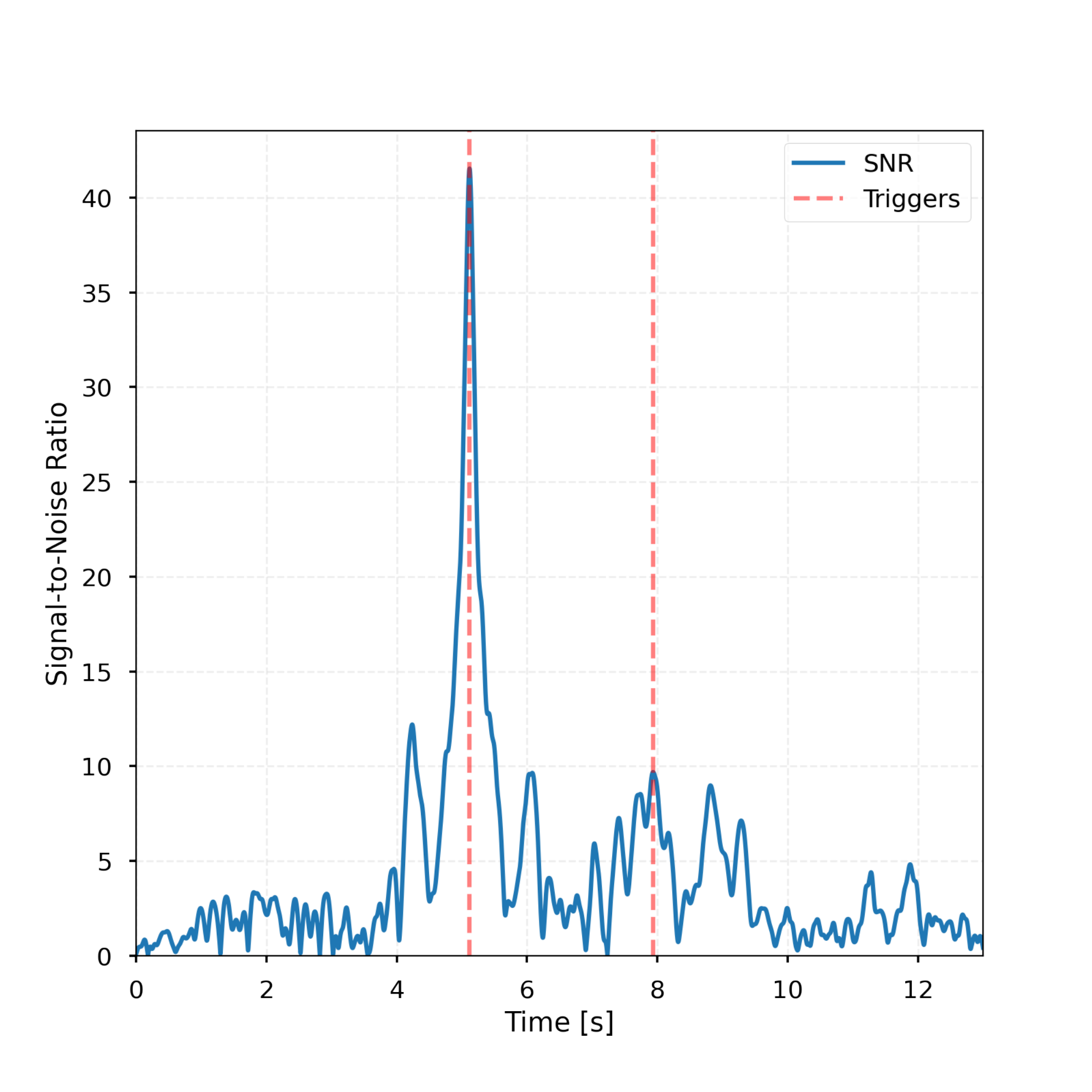}
      \end{minipage}
      \begin{minipage}[t]{1.0\linewidth}
        \includegraphics[width=0.49\linewidth]{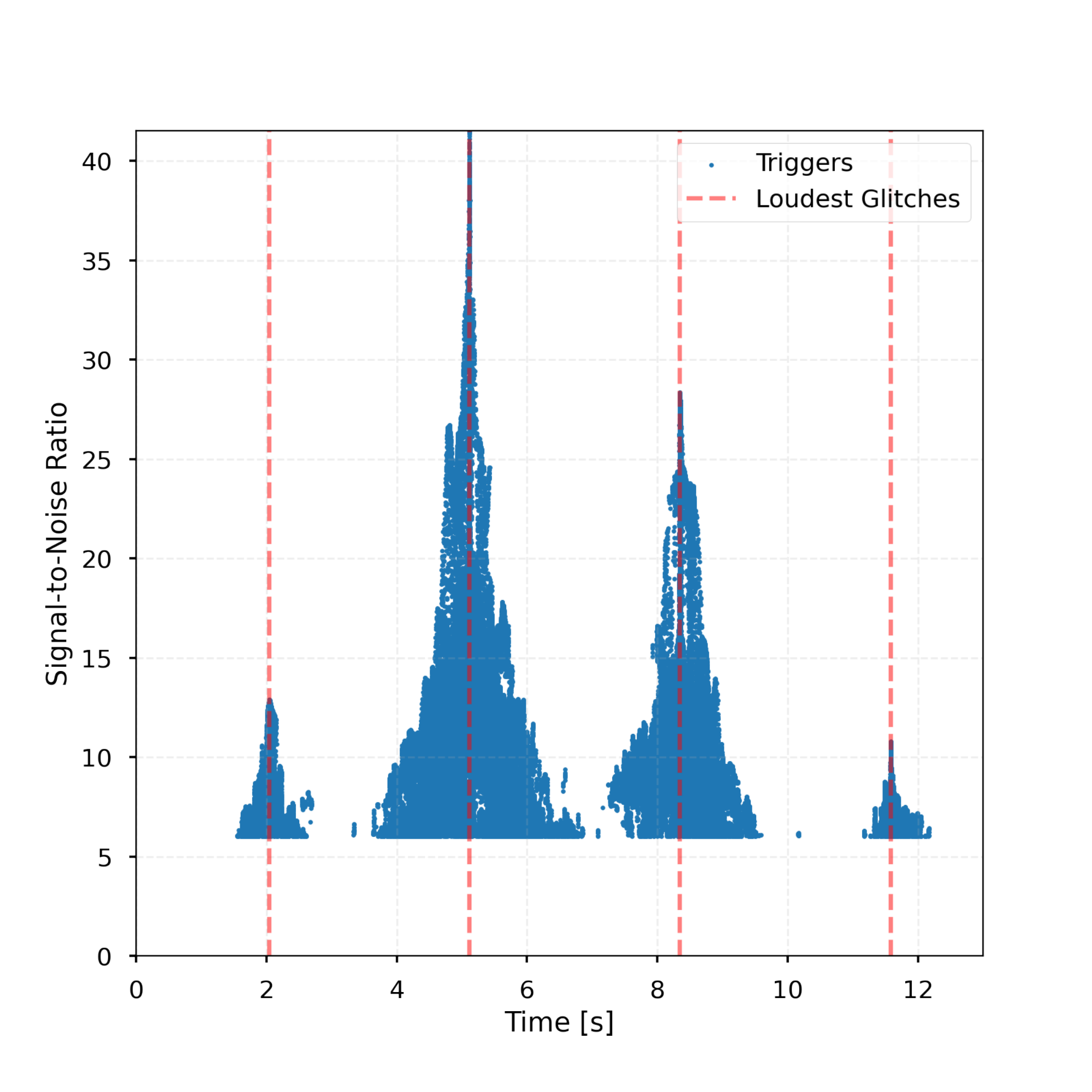}
        \hspace{0.01\linewidth}
        \includegraphics[width=0.49\linewidth]{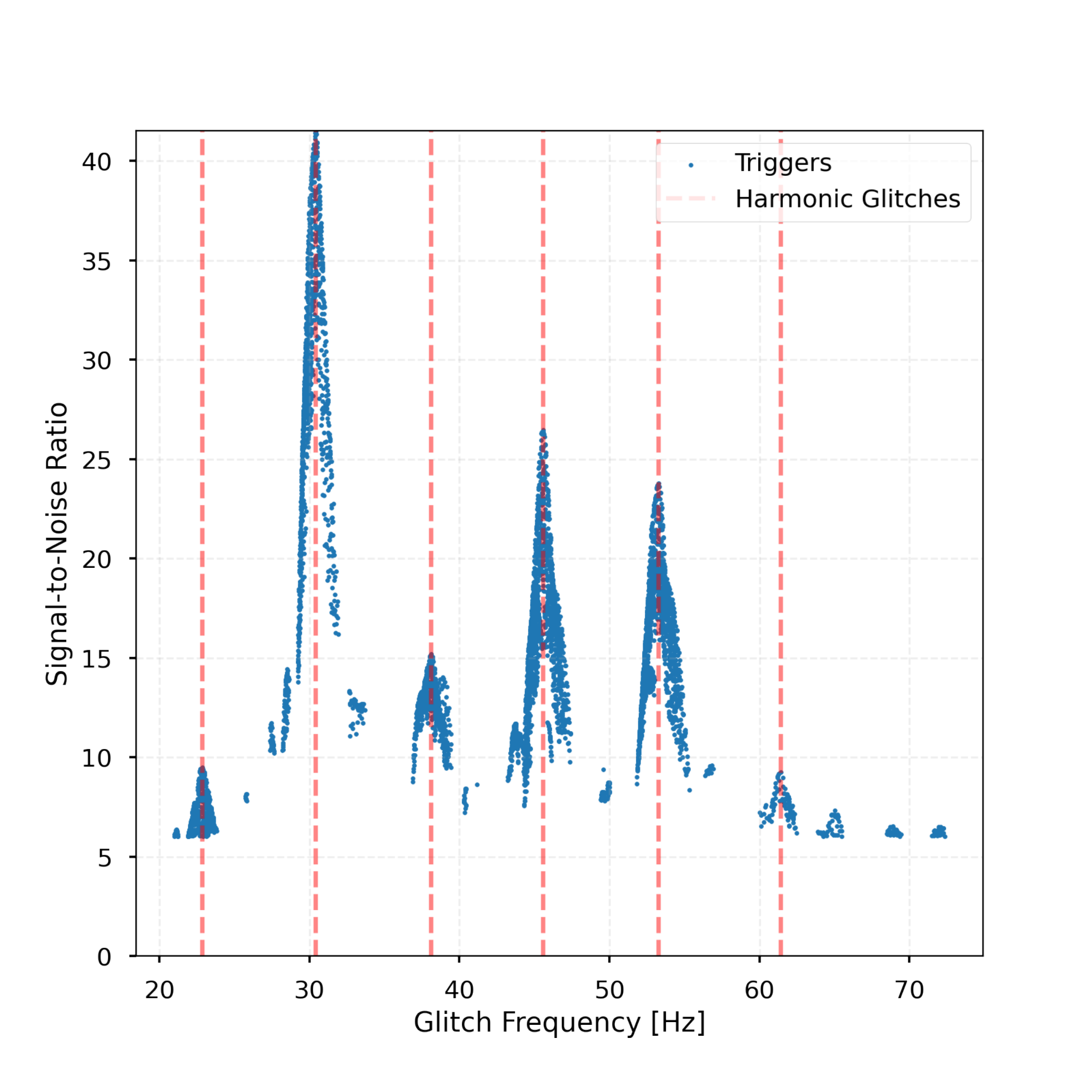}
      \end{minipage}
    \caption{The process for identifying all the \scl{} glitches in a period of \gw{} data. The red overlay in the \gw{} data used in this example (top left) indicates the highest re-weighted signal-to-noise trigger found, the dashed vertical lines represent the time slice window around this trigger. The re-weighted signal-to-noise ratio time series resulting from the matched filter of this trigger's template and the data (top right) is clustered in time to identify the triggers found above a signal-to-noise threshold of $8$, indicated by red vertical dashed lines. All the triggers from all of the templates in the template bank are then clustered in time (bottom left) to identify the highest re-weighted signal-to-noise glitches in the data, indicated by the orange vertical dashes lines. The triggers found within the time slice window, with a similar \emph{time period} value---within $\pm 10\%$---of the highest re-weighted signal-to-noise ratio trigger (bottom right) are clustered by their \emph{glitch frequency} value to find the harmonic glitches at the same time, indicated again by red dashed lines.}
    \label{fig:clustering_story}
\end{figure}

Our matched filtering process retains ``triggers'' (potential \scl{} glitches) wherever the re-weighted signal-to-noise time series is larger than 8. We retain no more than one trigger within a window size equal to half the \emph{time period} of the template used to produce the re-weighted signal-to-noise time series and only store triggers at local maxima. A re-weighted signal-to-noise time series with multiple peaks and identified triggers can be seen in the top right panel in figure \ref{fig:clustering_story}.

After matched filtering all the templates against the data we will recover multiple triggers for any potential \scl{} glitch, as we might expect to independently identify peaks in multiple templates around the true values of the glitch. We therefore collect all of the triggers generated by the template bank and cluster these in time, using a window of half of the shortest duration template - $0.9$ seconds. This will result in a list of triggers corresponding to the highest re-weighted signal-to-noise ratios, where each trigger should correspond to a unique \scl{} glitch. The bottom left panel in figure \ref{fig:clustering_story} shows an example of the triggers found by the search and the highest re-weighted signal-to-noise ratio triggers found by clustering.

However, we also expect to see instances of harmonic glitches which are produced by the same scattering surface and so share the same \emph{time period} and have \emph{glitch frequency} values equal to a multiple of the lowest frequency glitch in the harmonic stack~\cite{TAccadia}. We investigate each trigger in the list, searching for harmonic glitches occurring at the same time. We use the first list of triggers identified by all templates across the bank and filter by those that occur within $\pm0.05$ seconds of the \emph{center time} of the trigger we are investigating, an example of this window can be seen in the top left panel of figure \ref{fig:clustering_story}. We then filter the triggers again, keeping only those with an associated \emph{time period} value within $\pm 10 \%$ of the trigger's \emph{time period}. Finally we cluster these remaining triggers by their associated \emph{glitch frequency} using a window size of $4$Hz, a lower limit on the frequency separation of harmonic glitches, the bottom right panel in figure \ref{fig:clustering_story} shows the identification of harmonic glitches for the overlaid \scl{} glitch in the top left panel of figure \ref{fig:clustering_story}.

\subsection{Hierarchical subtraction to find parameter values}

\begin{figure}
     \centering
     \begin{minipage}[t]{1.0\linewidth}
        \includegraphics[width=0.49\linewidth]{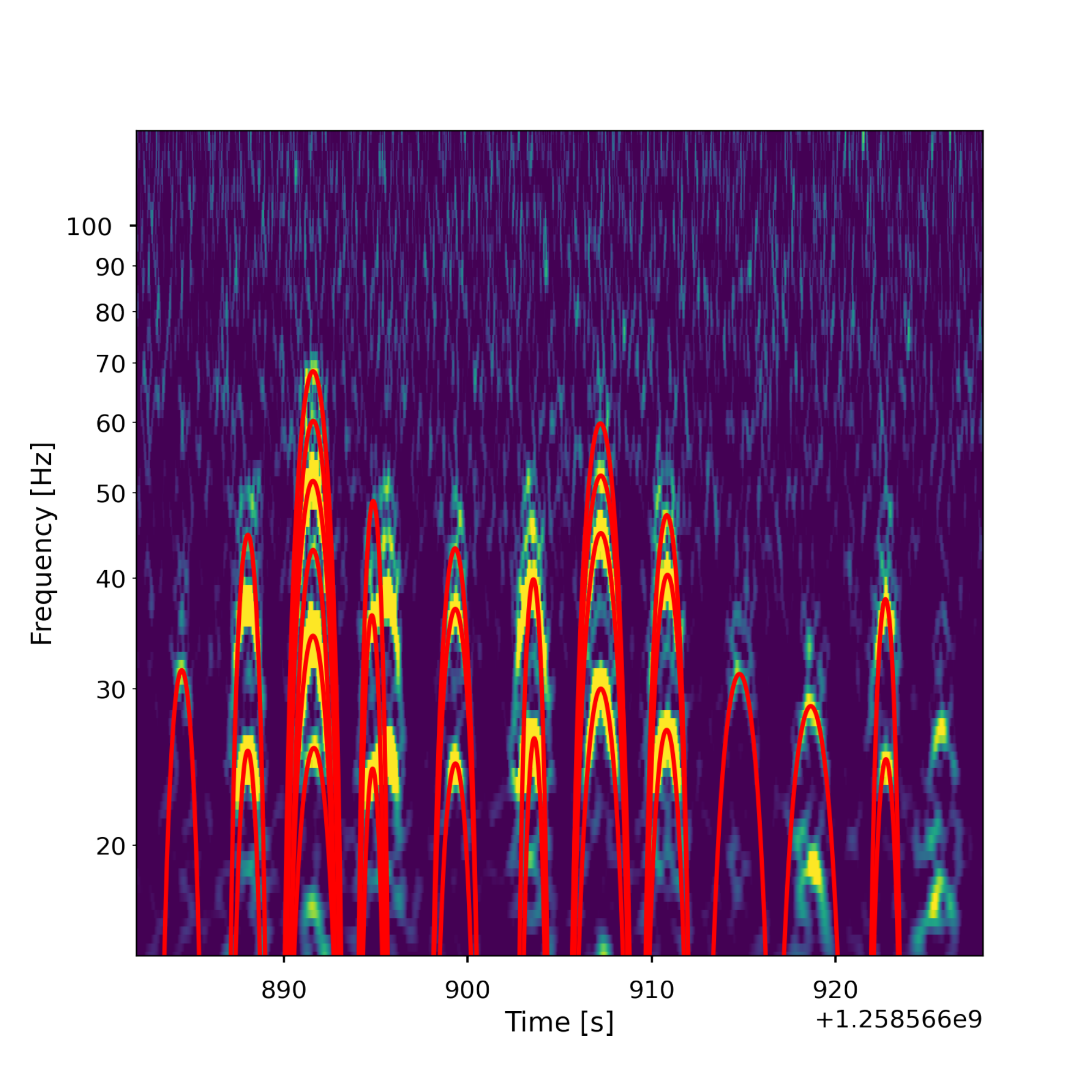}
        \hspace{0.02\linewidth}
        \includegraphics[width=0.49\linewidth]{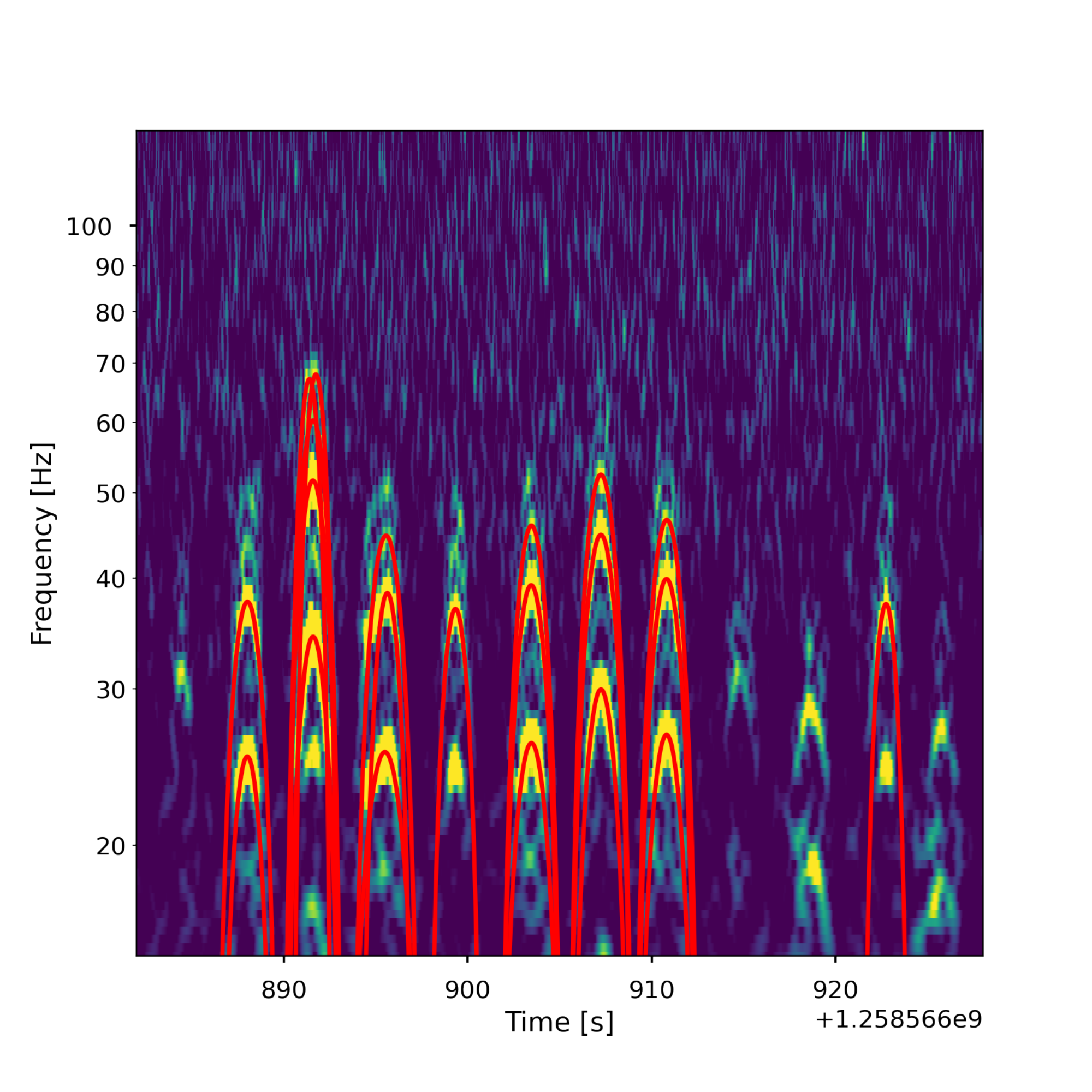}
     \end{minipage}
         \caption{LIGO-Hanford data from 2019-11-23 17:54:22 - 2019-11-23 17:55:12 containing \scl{} glitches which have been identified by the ArchEnemy search (left), there is a misalignment in the template found for a number of glitches in this period of data and some missed glitches. \Scl{} glitches remaining after running the hierarchical subtraction search (right) for the same period of data, we have missed more \scl{} glitches however misalignments have been removed. The highest harmonic at approximately $892$ seconds has been incorrectly split into two separate templates.}
    \label{fig:overlay_goods}
\end{figure}

We now have a list of identified \scl{} glitches and their parameter values, however, these might not be fully accurate when there are many glitches close in time and frequency, as illustrated in figure~\ref{fig:overlay_goods}. It is important that the parameters we find match well with the glitches in the data to remove as much power as possible.

To better identify the parameter values of the \scl{} glitches, we perform a hierarchical procedure using information about the glitches we have found so far. Firstly, we create new segments of time which we know contain \scl{} glitches, taking a window of $8$ seconds on either side of each previously identified glitch, if two glitch windows overlap they are combined into the same segment.

For each segment we then create a reduced template bank, consisting of templates ``close'' to the \scl{} glitches previously identified in the segment. We take the smallest and largest \emph{time period} and \emph{glitch frequency} glitches in the segment and bound the retained templates by these values with $\pm 0.25$ seconds on the \emph{time period} and $\pm 1$Hz on the \emph{glitch frequency}. For each data segment the reduced template bank is matched filtered with the data, the maximum re-weighted SNR value is found and the corresponding glitch is subtracted. We then matched filter \emph{again} and remove the next largest re-weighted SNR template. This process is repeated until no templates, when matched filtered with the data, produce any re-weighted SNR values above the SNR limit of $8$. This method of hierarchical subtraction produces our final list of \scl{} glitches.

A further benefit of using these shorter data segments is that we are estimating the PSD of the data using only a short period of data close to the \scl{} glitches being removed. This protects us from a rapidly changing PSD in non-stationary data, which might cause Gaussian noise to be identified with larger SNR in the periods where the PSD is larger. This can be resolved by including the variation in the power spectral density as an additional statistic in the re-ranking of triggers, which has been done for compact binary coalescence \gw{} searches in~\cite{Mozzon_2020}.

\begin{figure}
     \centering
     \begin{minipage}[t]{1.0\linewidth}
        \includegraphics[width=0.49\linewidth]{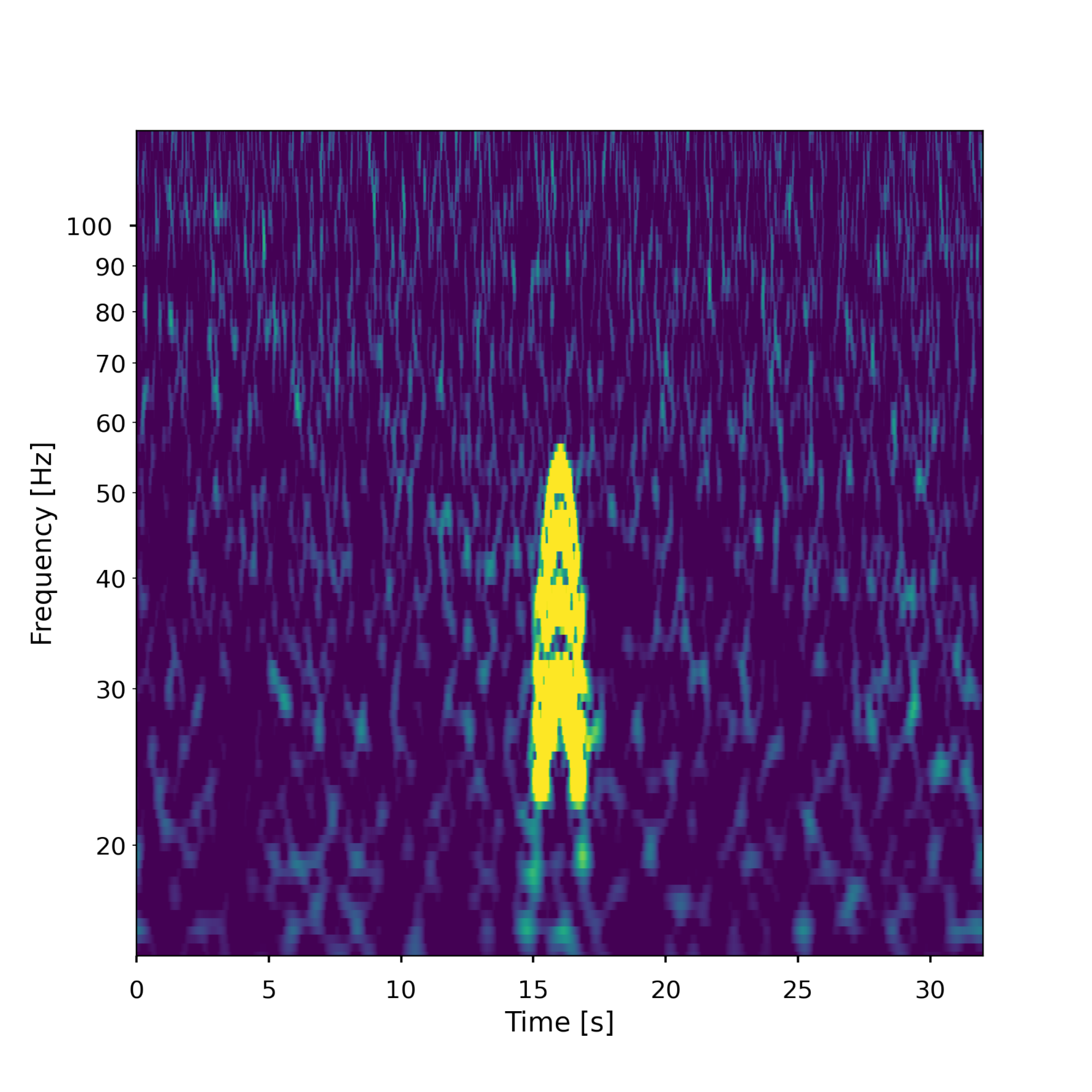}
        \hspace{0.02\linewidth}
        \includegraphics[width=0.49\linewidth]{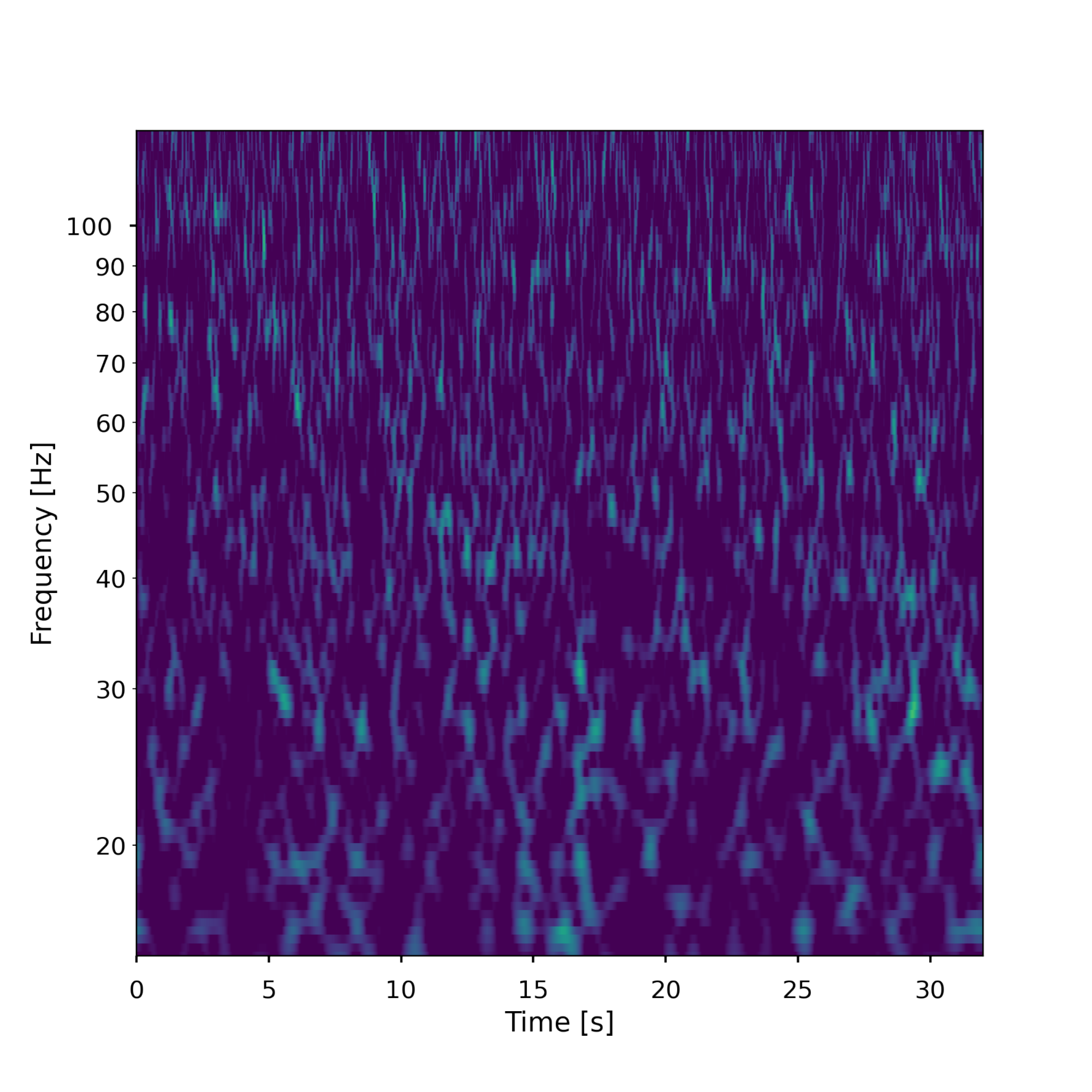}
     \end{minipage}
         \caption{Data containing an injected stack of harmonic \scl{} glitches (left) and the corresponding data found when running the hierarchical subtraction search and subtracting the identified \scl{} glitches from the data (right).}
    \label{fig:injected_glitches}
\end{figure}

We demonstrate the hierarchical subtraction step on an amount of data which contains four injected \scl{} glitches in a single harmonic stack, this can be seen in figure \ref{fig:injected_glitches}. As shown, the \scl{} glitches are found and subtracted from the data leaving behind a cleaned segment of \gw{} data with no excess noise. Figure~\ref{fig:overlay_goods} shows the identified \scl{} glitch triggers before and after performing the hierarchical subtraction step on a stretch of data containing real \scl{} glitches. We identify more triggers prior to performing hierarchical subtraction, however there are more errant mismatches between \scl{} glitches and the overlaid templates. By performing the hierarchical subtraction, we more accurately identify \scl{} glitches, however, we miss some glitches that were previously identified. There is still some imperfection in this process and we do not refine the method further in this work, but highlight this as useful direction for future studies in removing \scl{} glitches. 

\subsection{Identified \scl{} glitches}

The methodology described in previous sections is implemented in our ``ArchEnemy'' pipeline, which is capable of searching for \scl{} glitches in \gw{} data using a pre-generated bank of glitch templates. We use ArchEnemy to analyse the aforementioned data, which produced a list of $2749$ \scl{} glitches in data from the LIGO-Hanford observatory and $1306$ from the LIGO-Livingston observatory.

\begin{figure}
       \centering
    \includegraphics[width=0.7\linewidth]{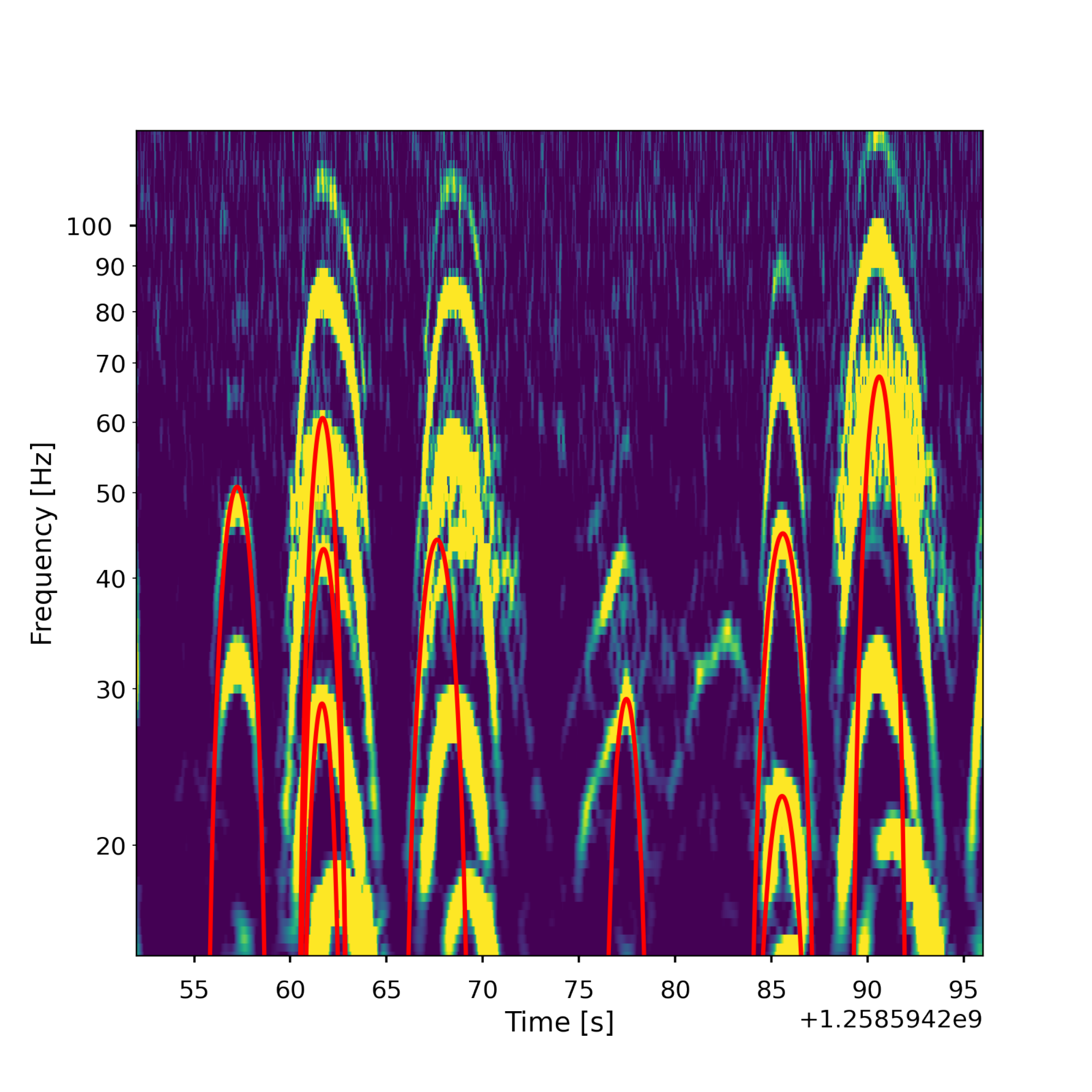}
    \caption{LIGO-Livingston data from 2019-11-24 01:30:32 - 2019-11-24 01:31:20 containing a very large number of \scl{} glitches at multiple times and frequencies over-plotted with the \scl{} glitches identified by the ArchEnemy search. Very few overlays match well onto \scl{} glitches. The template bank used in this search terminates at $80$ Hz and so the \scl{} glitches located above this value will not be correctly identified. There are also asymmetric \scl{} glitches located which will not be identified correctly by our search which assumes symmetry in the \scl{} glitch.}
    \label{fig:overlay_bads}
\end{figure}

The number of \scl{} glitches found by the ArchEnemy pipeline can be compared to Gravity Spy for the same period of time. Gravity Spy finds $2731$ and $1396$ for LIGO-Hanford and LIGO-Livingston respectively~\cite{gravityspy}. There will be a difference in the number of glitches found by ArchEnemy and Gravity Spy for at least two reasons: Gravity Spy treats an entire stack of harmonic glitches as a single \scl{} glitch whereas ArchEnemy will identify each glitch as a separate occurrence. Gravity Spy can also identify \scl{} glitches which are not symmetric and fall outside our template bank, for example, the \scl{} glitches shown in figure~\ref{fig:overlay_bads}.

Figure~\ref{fig:overlay_goods} is an example of the results of the ArchEnemy pipeline and how well it has identified \scl{} glitches in a period of data. A majority of the glitches have been identified with the correct parameter values and even in cases where the chosen template is not visually perfect, there is a good match between the template and the identified power in the data, particularly in the case of slightly asymmetric glitches. Figure~\ref{fig:overlay_bads} demonstrates a period of time where the ArchEnemy pipeline has not fitted well the \scl{} glitches in the data. The glitches at this time are improperly fit by the templates due to asymmetry of the morphology of the glitches and because some of the glitches are outside of our template bank parameter range. However, we note that this is a very extreme period of \scl{} glitching and immediately after this time the detector data is no longer flagged as ``suitable for analysis''.

We have demonstrated the ArchEnemy pipeline on a stretch of data from O3 and have identified and characterized a list of \scl{} glitches, which could be removed from the data. We do note that there are cases where the identification has not worked well, but we expect that subtracting our list of glitches from the data will reduce their effect on the \gw{} search. In the next section we will demonstrate this by quantifying sensitivity with the PyCBC pipeline.

\subsection{Safety of \scl{} identification}
\label{ssec:injsafety}

The data we have searched through contains no previously identified \gw{} signals~\cite{gwtc3}. However, there is a risk that the ArchEnemy search would identify real \gw{} signals as \scl{} glitches. To assess this possibility we simulate and add a large number of \gw{} signals into the data and assess whether any signals are misidentified.

To do this, we use three separate sets of simulated \gw{} signals (or ``injection sets''), one for BBHs, another for binary neutron stars (BNSs) and a third for neutron star black hole (NSBH) systems. We use the same simulations as the LVK search of this data, detailed in the appendix of~\cite{gwtc3}. Each injection set consists of $6200$ simulated signals spaced between $82$ and $120$ seconds apart. We treat these injection sets exactly the same as for the injection-less data, adding the simulations to the data, and then running ArchEnemy to produce a list of \scl{} glitches for each injection set.

To determine whether we have misidentified any \gw{} injections as \scl{} glitches we look for glitches we have found within the overlapping frequency band of \gw{} signals and our \scl{} glitch template bank. This corresponds to approximately $15$ second before merger time for the injections. The simulated signals occur every $\sim100$ seconds so we expect to see glitches within this $15$ second window, therefore, we also require that there must be more triggers identified in the \scl{} glitch search \emph{with} injections when compared to the search \emph{without} injections within the window. The details of the number of \gw{} injections with overlapping \scl{} triggers can be seen in table~\ref{tab:coincident_triggers}.

\begin{table}[tb]
\caption{\label{tab:coincident_triggers}For both interferometers and all $3$ injection sets we identify the number of injections which are found to have \scl{} glitches identified within $15$ seconds of merger time (``Injections with Coincident Triggers''), along with the number of \scl{} glitches found within this window for these injections (``Scattered-Light Coincident Triggers''). We investigated each of these injections and recorded the number which actually had \scl{} glitches identified due to the injected \gw{} signal (``Actual Overlapped Injections'').}
\footnotesize
\begin{tabular}{ccccc}
\br
&    & Injections with& Scattered-Light & Actual Overlapped \\
Interferometer & Injection Set & Coincident Triggers& Coincident Triggers&Injections \\
\mr
H1 & BBH           & 20 & 45 & 1\\
   & BNS           & 23 & 50 & 2\\
   & NSBH          & 38 & 73 & 7\\
L1 & BBH           & 13 & 21 & 2\\
   & BNS           & 18 & 30 & 2\\
   & NSBH          & 35 & 56 & 16\\
\br
\end{tabular}

\end{table}

A \scl{} glitch will be identified close to a \gw{} signal in two cases: the ArchEnemy search is misidentifying the \gw{} signal as a glitch \emph{or} the simulated signal was added close to actual glitches and a change in the data has meant a different number of glitches has been identified. The presence of real \scl{} glitches means we might miss a \gw{} signal, therefore, we \emph{do} want to find and subtract glitches close to \gw{} signals, but we do not want to subtract power from the \gw{} signal itself. The \scl{} glitch $\chi^{2}$ test was designed to prevent the matching of \scl{} glitch templates on other causes of excess power, however, these results show it is not perfect.

\begin{figure}
  \centering
  \begin{minipage}[t]{1.0\linewidth}
    \includegraphics[width=0.49\linewidth]{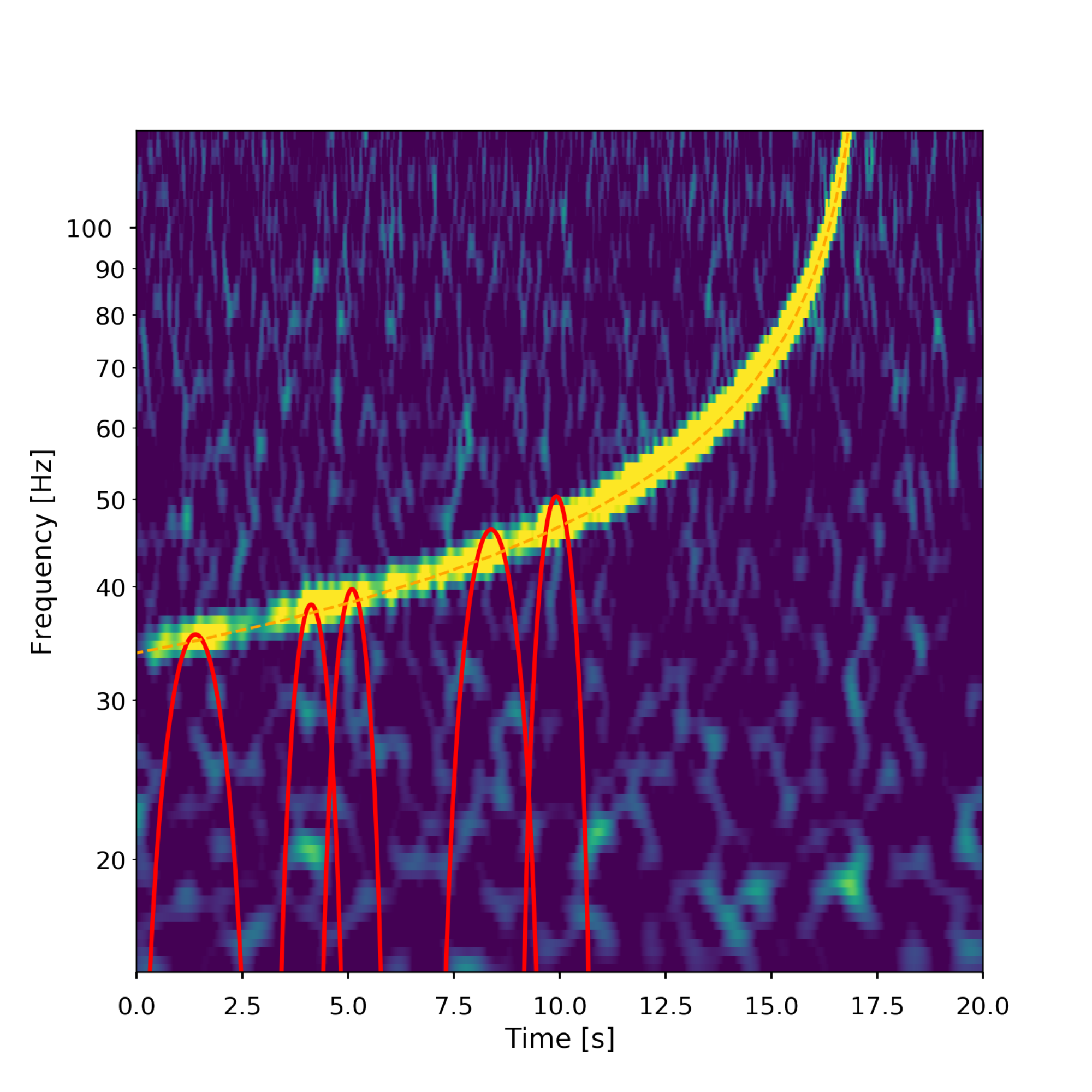}
    \hspace{0.02\linewidth}
    \includegraphics[width=0.49\linewidth]{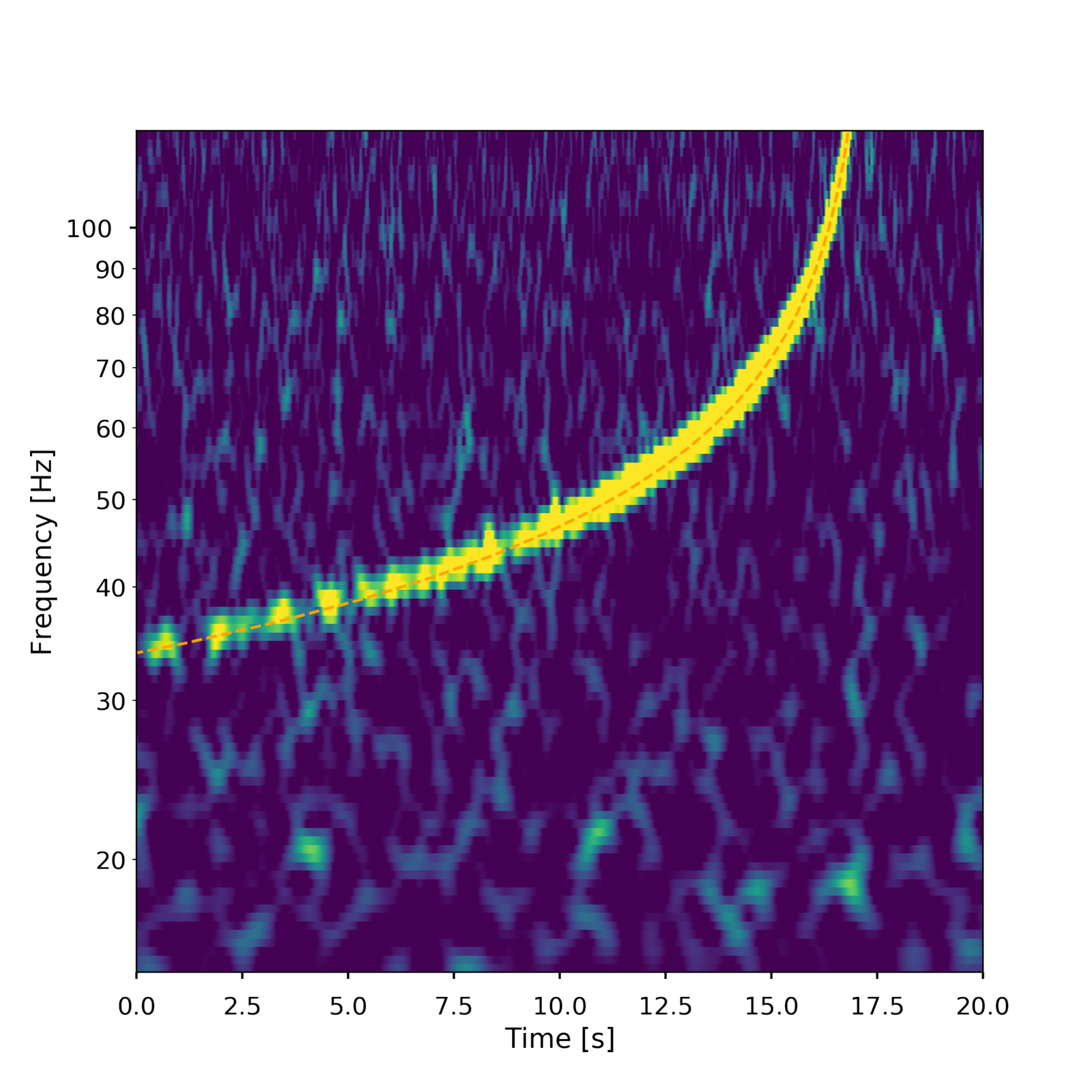}
  \end{minipage}
    \caption{An injected binary neutron star compact binary coalescence \gw{} signal, with the \scl{} glitches identified by the ArchEnemy search pipeline overlayed in red (left). The same injected signal but with the \scl{} glitches removed from the data (right). It can be seen that power is removed from the signal track and also there is an amount of power added to the data above the track.}
    \label{fig:loud_inj}
\end{figure}

We investigate each injection with coincident \scl{} triggers, seeing how many had misidentified \scl{} glitches on the inspiral of the \gw{} signal, this number can be seen in the column ``Actual Overlapped Injections'' in table~\ref{tab:coincident_triggers}. We have included an example of the matching of \scl{} glitches onto \gw{} injections in figure~\ref{fig:loud_inj}, the right panel shows the \gw{} data post glitch subtraction where it can be seen there is a portion of the power being subtracted from the signal. Although power is being removed from the signal, the \gw{} injection is still found by the search for \gws{}, which we will describe later. For the cases that we have investigated, we note that the behaviour shown in figure~\ref{fig:loud_inj} only happens for signals that have very large signal-to-noise ratio, and are therefore unphysically close to us. A similar effect is observed with the ``autogating'' process, described in~\cite{pycbc}, which prevents the detection of these loud signals. In contrast to the ``autogating'' though, signals like that illustrated in figure~\ref{fig:loud_inj} are still identified as \gw{} signals by the PyCBC search after \scl{} glitch removal.

\section{\label{sec:results}Assessing sensitivity gain from removing \scl{} glitches}

We now assess whether removing our identified list of \scl{} glitches results in a sensitivity gain when searching for compact binary mergers. We do this by comparing the results from the offline PyCBC search on the original data, to the results of the same search but analysing data where the glitches have been removed.

\subsection{Comparing search results with and without glitch subtraction}

The PyCBC pipeline is able to assess significance of potential compact binary mergers in a given stretch of data, and does the same with a set of simulated signals. This significance is quoted in terms of a ``false-alarm rate'', which denotes how often we would expect to see a non-astrophysical event at least as significant as the coincident trigger being considered. In this work we assess sensitivity at a false-alarm threshold of 2 background events every year.

The data we have searched over contained no previously found \gw{} signals~\cite{gwtc3} and our search after subtracting \scl{} glitches identified no new \gw{} signals. While the search hasn't found any \gws{}, we can still measure the improvement in the sensitivity of the detectors by comparing the number of simulated signals identified with a false-alarm rate below 2 per year for each injection set (described in section~\ref{ssec:injsafety}) with and without removing glitches from the data. Table \ref{tab:found_injs} shows the number of injections found for all injection sets and both searches.
\begin{table}[tb]
\centering
\caption{\label{tab:found_injs}The number of injections found by each search with a false-alarm rate less than 2 per year alongside the number of newly-found and newly-missed injections, those found by the glitch-subtracted and not the original search and vice versa. We also show the sensitivity ratio of the glitch-subtracted search and original search for each injection set.} 
\footnotesize
\begin{tabular}{@{}cccccc}
\br
Injection & Original  & Glitch- & Sensitivity & Newly & Newly \\
Type & Search & Subtracted & ratio & Found & Missed \\
\mr
BBH           & 1215 & 1222 & 1.01 & 10 & 3\\
BNS           & 1315 & 1315 & 1.00 & 5 & 5 \\
NSBH          & 1260 & 1265 & 1.00 & 8 & 3 \\
\br
\end{tabular}

\end{table}

We compare the number of injections found by both searches but also look at the \gw{} injections found by the original search and missed by the glitch-subtracted search and vice versa, this information can be seen in table~\ref{tab:found_injs}. Considering signals found by the original search and missed by the glitch-subtracted search there are $3$ binary black hole injections with false-alarm rates in the original search ranging from $0.5 - 0.3$ per year, one of which had a glitch removed approximately $9$ seconds after the injection, there are $5$ newly-missed binary neutron star injections with false-alarm rates ranging from $0.5 - 0.056$ per year, two of the five binary neutron star injections had glitches removed within $60$ seconds of the injection, and there are $3$ newly-missed neutron star black hole injections with false-alarm rates ranging from $0.5 - 0.086$, one had glitches removed within $60$ seconds of the injection. The other newly-missed injections showed no \scl{} glitches within a $20$ second window for binary black hole injections and a $60$ second window for binary neutron star and neutron star black hole injections.

\begin{figure}
  \centering
  \begin{minipage}[t]{1.0\linewidth}
    \includegraphics[width=0.49\linewidth]{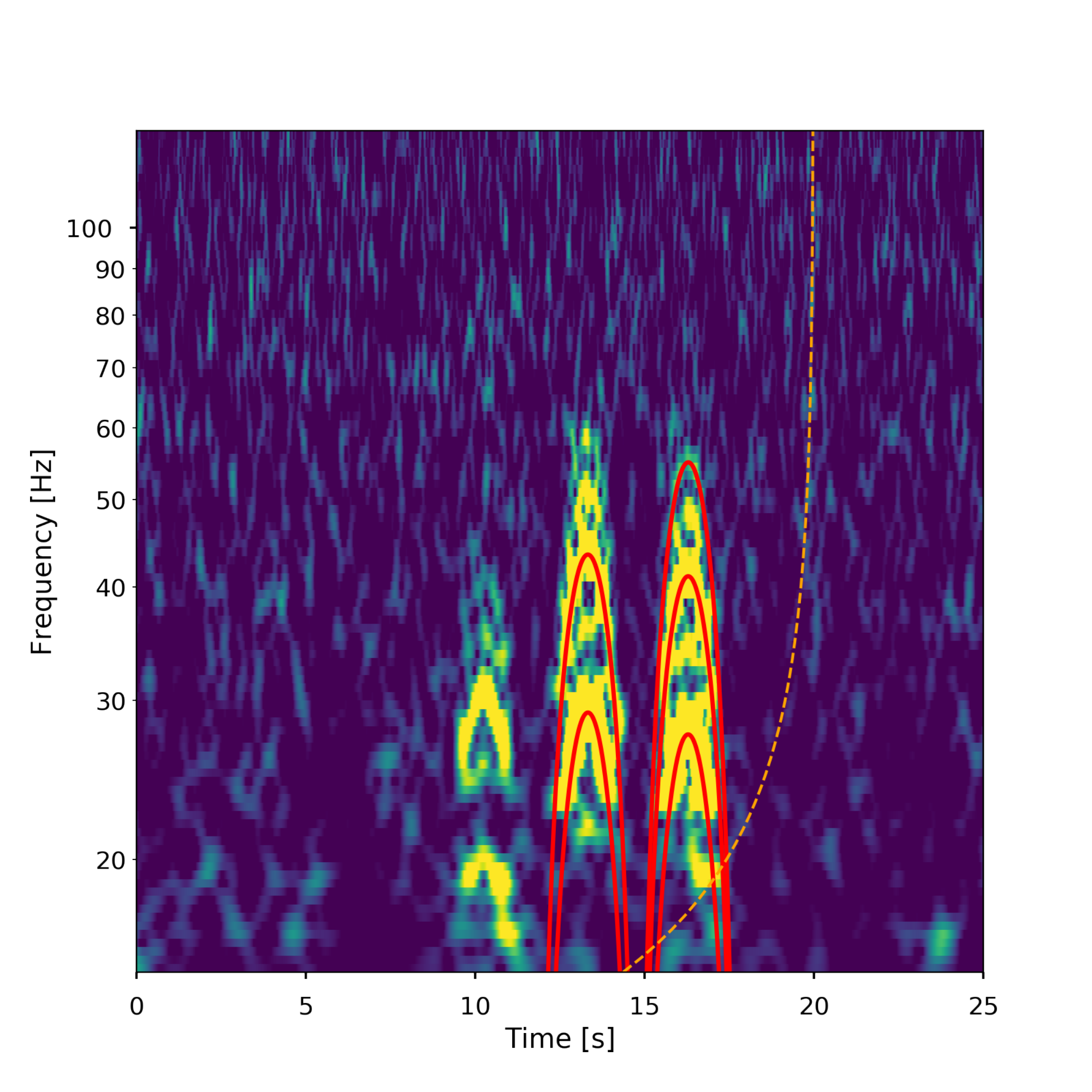}
    \hspace{0.02\linewidth}
    \includegraphics[width=0.49\linewidth]{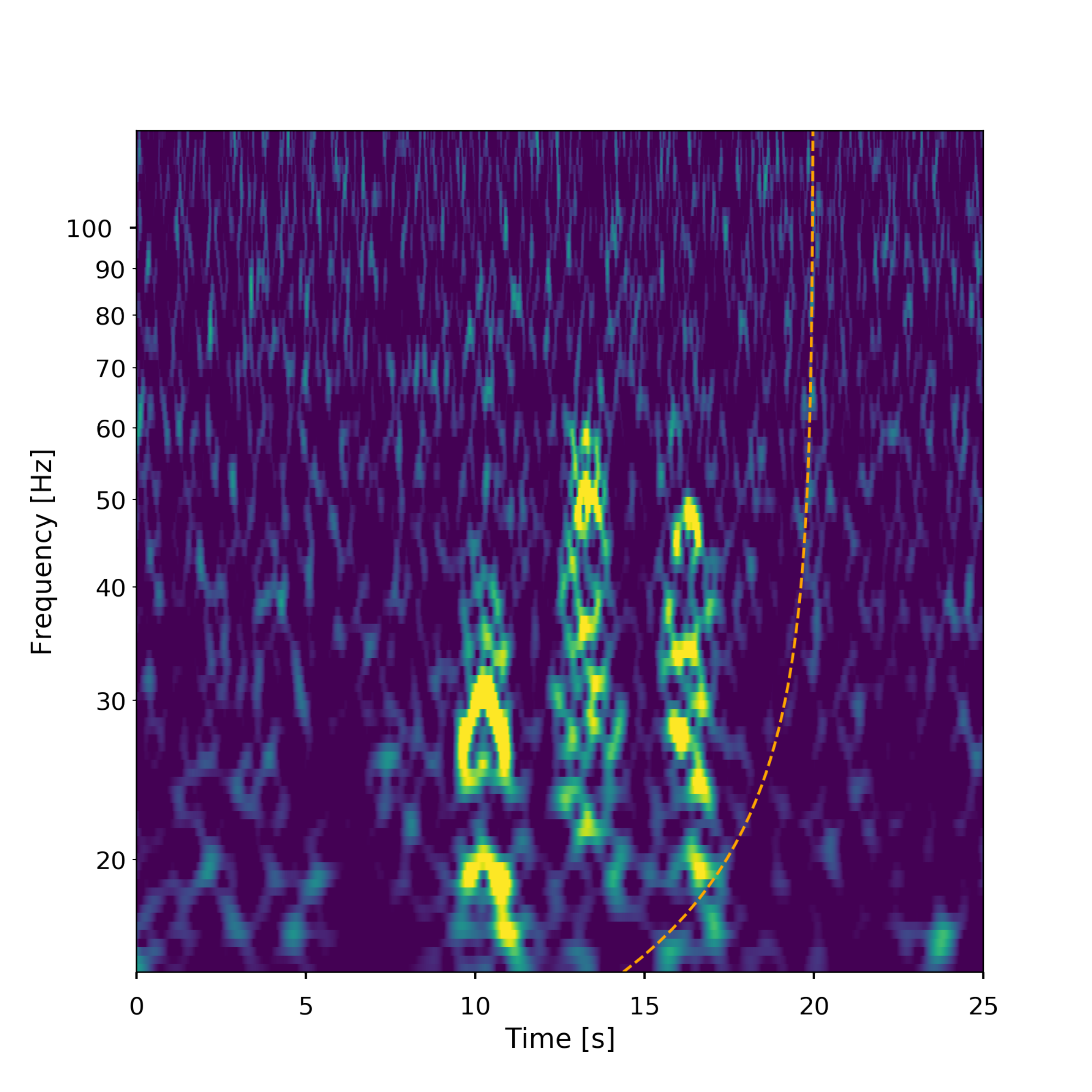}
  \end{minipage}
  \begin{minipage}[t]{1.0\linewidth}
    \includegraphics[width=0.49\linewidth]{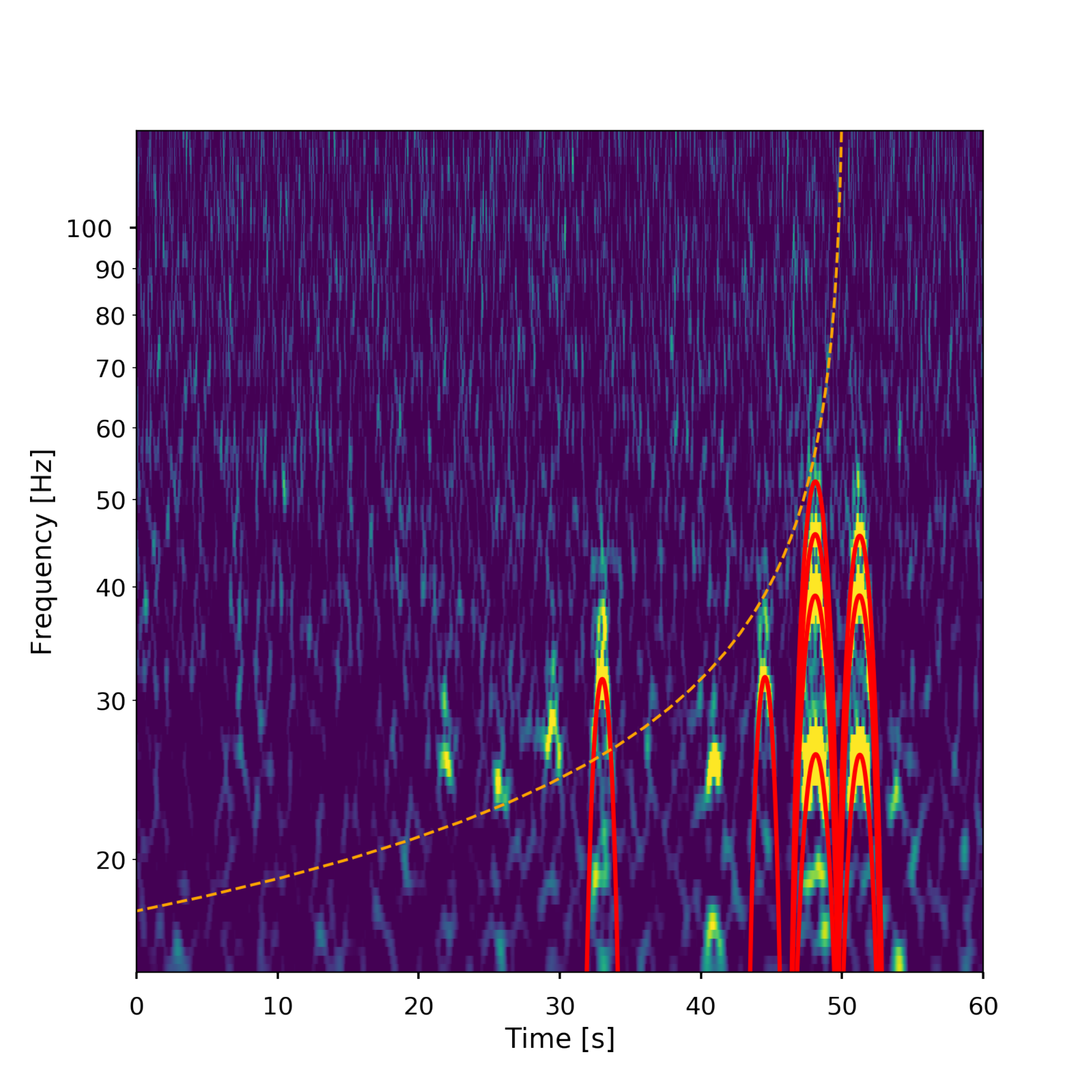}
    \hspace{0.02\linewidth}
    \includegraphics[width=0.49\linewidth]{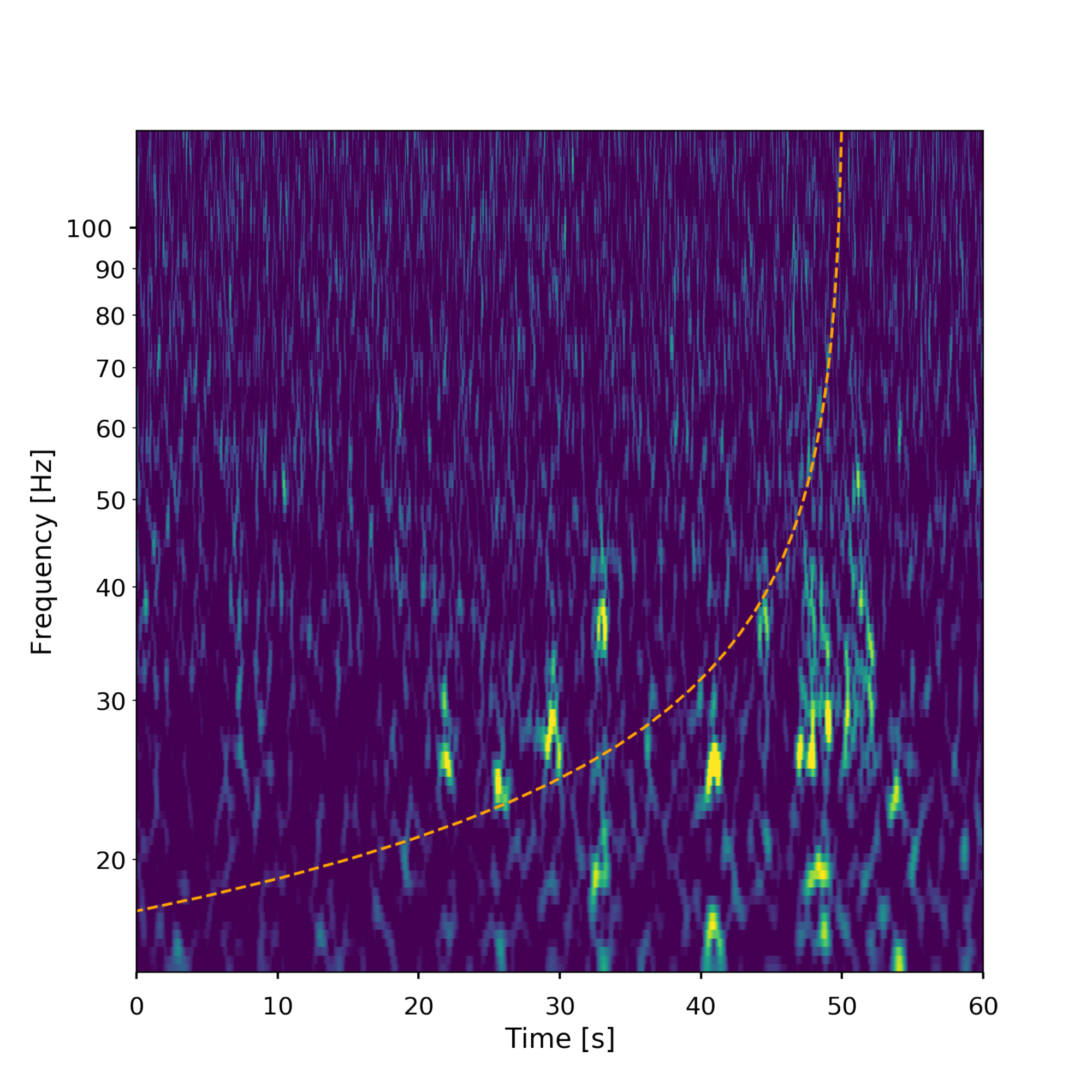}
  \end{minipage}
    \caption{Two examples of \gw{} injections found by the glitch-subtracted search for \gws{} which were not found by the original \gw{} search due to the presence of \scl{} glitches at the same time as the \gw{} inspiral. Top left: A binary black hole injection with a false-alarm rate of 1 per $8643.73$ years is shown alongside the \scl{} glitches found by the ArchEnemy search and subtracted from the data prior to performing the glitch-subtracted PyCBC search for \gws{} (top right). Bottom left: A neutron star black hole injection with a false-alarm rate of 1 per $9961.55$ years and the \scl{} glitches found by the ArchEnemy search and subtracted from the data prior to performing the glitch-subtracted PyCBC search for \gws{} (bottom right).}
    \label{fig:ae_found}
\end{figure}

The glitch-subtracted search identifies $10$ additional binary black hole injections, the most significant of which have false-alarm rates of 1 per $190.50$, 1 per $7633.84$ and 1 per $8643.73$ years. We illustrate the last of these in figure~\ref{fig:ae_found} (top). $5$ extra binary neutron star injections were found, with false-alarm rates from $0.5 - 0.19$ per year and $8$ neutron star black hole injections were found, where the false-alarm rate of the most significant is 1 per $9961.55$ years. This injection can also be seen in figure~\ref{fig:ae_found} (bottom). We find $6$ of the $10$ binary black hole injections have \scl{} glitches within a $20$ second window of the injection, $3$ of $5$ binary neutron star injections have \scl{} glitches within a $60$ second window of the injection and, $6$ of the $8$ neutron star black hole injections have \scl{} glitches within a $60$ second window of the injection. We provide more details about the newly-found and newly-missed injections in~\ref{sec:apdx_injections_table}

To quantify the sensitivity of the search we calculate the sensitive volume in which we can observe \gw{} signals. To calculate the sensitive volume we measure the detection efficiency of different distance bins taken from the injection sets and then multiply the efficiencies by the volume enclosed by the distance bins, these volumes are then summed to find the total volume the search is sensitive to~\cite{rw_snr_eq}. We are then able to calculate the ratio in sensitivities between the glitch-subtracted \gw{} search and the original \gw{} search,  revealing the improvement that subtracting \scl{} glitches has made.

\begin{figure}
     \centering
     \includegraphics[width=0.7\textwidth]{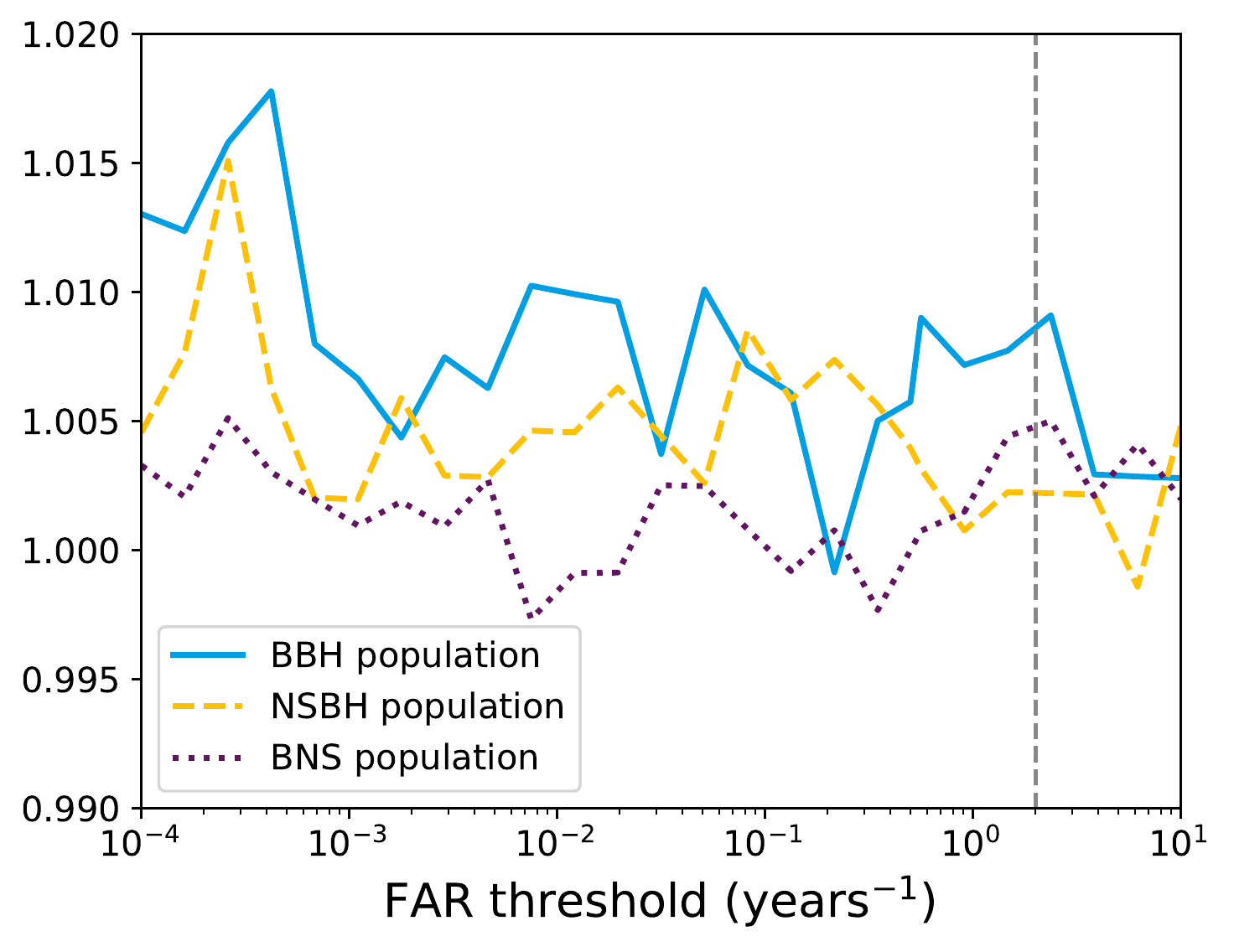}
     \caption{The ratio of the sensitive volume-time of the glitch-subtracted search and the original \gw{} search. The grey dashed line indicates a false-alarm rate of $2$ per year which is our threshold and the point at which we measure any sensitivity improvements of the glitch-subtracted search for each of the three \gw{} injection sets. }
     \label{fig:allinj_vt_ratio}
\end{figure}

Figure~\ref{fig:allinj_vt_ratio} displays the ratio of the sensitive volume measured for the glitch-subtracted \gw{} search and the original PyCBC \gw{} search, across different false-alarm rate values, we quote our sensitivity ratios at a false-alarm rate value of 2 per year. The same set of injected signals was used for both \gw{} searches and therefore a direct comparison of search sensitivities can be made via this ratio. Disappointingly, the measured sensitivity improvement is small in the results we obtain. For the binary black hole injections we measure a sensitivity ratio at a 2 per year false-alarm rate of $1.01$, for binary neutron stars $1.00$ and neutron-star--black-holes $1.00$. 

The statistical significance of the sensitivity increase we report for the binary black hole injection set can be found by investigating the null hypothesis of seeing the same $1\%$ increase under the assumption that the subtraction of \scl{} glitches does not actually increase sensitivity. When performing this analysis we find that our result is not statistically significant at the 95\% confidence interval --i.e. there is a 5.24\% chance that we would measure such an increase in sensitivity at least as large as this under the null hypothesis-- (see~\ref{sec:apdx_stat_sig} for details). However, the marginal sensitivity increase would not justify repeating the \scl{} glitch search and glitch-subtracted \gw{} search on a larger injection set, instead more work is needed to better identify and remove \scl{} glitches while remaining safe in the presence of \gw{} signals.

\section{\label{sec:conclusion}Conclusion}

We have demonstrated a new method for modelling \scl{} glitches and identifying and characterizing these glitches in a period of \gw{} data. We have developed a \scl{} glitch specific $\chi^{2}$ test which can discriminate between \scl{} glitches, other types of glitches and \gw{} signals. We have searched through a representative stretch of \gw{} data known to contain \scl{} glitches, found thousands of these glitches and subtracted them from the \gw{} data prior to running a search for \gws{}. The results of this search include a small increase in the measured sensitivity of the \gw{} search for binary black hole \gw{} signals, and modest change to sensitivity for binary neutron star and neutron star black hole \gw{} signals.

We highlight that the task of accurately identifying and parameterizing \scl{} glitches in the data is not a trivial one, especially where there are repeated, and harmonic, glitches present in the data. We have developed a new $\chi^{2}$ test to reduce the number of false identifications of \scl{} glitches, but we do still see cases where we have misidentified other glitches, and even a small number of loud \gw{} signals, as caused by scattered light, and cases where we do not correctly identify, or parameterize, actual \scl{} glitches. Improving this identification process would be important in improving the efficacy of this process.

The possibility of using this model of \scl{} glitches as a bespoke application to \gw{} signals which are known to have coincident \scl{} glitches has been explored and implemented into Bilby~\cite{Bilby} to perform a parameter estimation of \scl{} glitches and removing these to produce glitch-free data~\cite{Udall2022}. The inclusion of the extra term from~\cite{MichalSub} within the model can help identify \scl{} glitches more accurately. Selectively subtracting glitches based on the presence of a \gw{} signal is possible by moving the glitch subtraction process inside of the \gw{} searches and including the results of other \gw{} discriminators~\cite{rw_snr_eq, McIsaac_2022} to determine the legitimacy of an ArchEnemy identified glitch.

The results of the application of the ArchEnemy search pipeline, the list of \scl{} glitches, can also be used in other applications. For example it could be used in the form of a veto~\cite{DetCharO2O3}, where we use knowledge of the presence of \scl{} glitches to down rank periods of time in \gw{} data. Additionally, we could use \scl{} glitches previously identified by tools such as Gravity Spy~\cite{GSpy2022} and target known \scl{} glitches with the ArchEnemy search pipeline.

As a final note, while we acknowledge that the sensitivity improvements that we have observed---$\sim 1\%$---are very modest, the concept of removing \scl{} glitches, or other identified glitch classes, from the data prior to matched filtering for compact binary mergers is one that we encourage others to explore further. An increase in the rate of events or the rate of \scl{} glitches in future observing runs will mean an increase in the number of affected events, such techniques offer a method for mitigating the effect that these glitches will have on the search, maximizing the number of observations that can be made.

\ack

We would like to thank Derek Davis for their useful comments on this work and the manuscript and also Rhiannon Udall for helpful discussion. AT would like to thank Laura Nuttall, Connor McIsaac, Connor Weaving and Ronaldas Macas for constant feedback, suggestions and debugging help during the development of this work. We would like to thank Thomas Dent for suggesting the name ``ArchEnemy''. AT was supported by the Science and Technology Facilities Council through the DISCnet Centre for Doctoral Training grant ST/V506977/1. GCD, IH and AL thank the STFC for support via the grants ST/T000333/1 and ST/V005715/1. The authors are grateful for computational resources provided by Cardiff University supported by STFC grant ST/I006285/1 and the LIGO Laboratory supported by the National Science Foundation Grants PHY-0757058 and PHY-0823459. This material is based upon work supported by NSF's LIGO Laboratory which is a major facility fully funded by the National Science Foundation. This research has made use of data or software obtained from the Gravitational Wave Open Science Center (gwosc.org), a service of LIGO Laboratory, the LIGO Scientific Collaboration, the Virgo Collaboration, and KAGRA. LIGO Laboratory and Advanced LIGO are funded by the United States National Science Foundation (NSF) as well as the Science and Technology Facilities Council (STFC) of the United Kingdom, the Max-Planck-Society (MPS), and the State of Niedersachsen/Germany for support of the construction of Advanced LIGO and construction and operation of the GEO600 detector. Additional support for Advanced LIGO was provided by the Australian Research Council. Virgo is funded, through the European Gravitational Observatory (EGO), by the French Centre National de Recherche Scientifique (CNRS), the Italian Istituto Nazionale di Fisica Nucleare (INFN) and the Dutch Nikhef, with contributions by institutions from Belgium, Germany, Greece, Hungary, Ireland, Japan, Monaco, Poland, Portugal, Spain. KAGRA is supported by Ministry of Education, Culture, Sports, Science and Technology (MEXT), Japan Society for the Promotion of Science (JSPS) in Japan; National Research Foundation (NRF) and Ministry of Science and ICT (MSIT) in Korea; Academia Sinica (AS) and National Science and Technology Council (NSTC) in Taiwan. This work carries LIGO document number P2200393. For the purpose of open access, the author has applied a Creative Commons Attribution (CC BY) licence to any Author Accepted Manuscript version arising. The data that support the findings of this study are openly available at the following URL: \href{https://github.com/ArthurTolley/ArchEnemy}{https://github.com/ArthurTolley/ArchEnemy}.

\rule{13.5cm}{0.01cm}
\bibliography{refs}
%
\appendix

\section{\label{sec:apdx_stat_sig}Statistical Significance}

We report a $1\%$ increase in the sensitivity for the binary black hole injection set in the glitch-subtracted \gw{} search (see section~\ref{sec:results}). We wish to determine the statistical significance of this result under the null hypothesis that subtracting \scl{} glitches prior to searching for \gws{} does not increase the sensitivity of the \gw{} search. The binary black hole injection set contains $6200$ injected signals, $10$ additional injections were found in the glitch-subtracted \gw{} search, $3$ additional injections were missed in our search, this gives us $13$ injections which have changed state.

First, we calculate the probability that an injection has been affected by the glitch removal, 
\begin{equation}
    p = \frac{13}{6200} = 0.21\% ,
\end{equation}
then we calculate the standard deviation, 
\begin{equation}
    \textrm{std} = \sqrt{n * p * (1 - p)} = 3.60 .
\end{equation}
Using the standard deviation we calculate the number of standard deviations our result deviates from the mean. We divide the number of positively changed (newly-found) injections by the standard deviation, 
\begin{equation}
    \textrm{standard deviations} = \frac{(10 - 3)}{3.60} = 1.94 .
\end{equation}
Under the assumption of no sensitivity increase caused by the subtraction of glitches, we measure our result of +7 newly-found \gw{} injections to lie $1.94$ standard deviations from an expected value of $0$ newly-found injections. The critical value of a 95\% confidence interval, that is to say there is a 1 in 20 chance of our null hypothesis being true, is $1.96$ meaning our result is within the 95\% confidence interval. We can describe this as there being a 5.24\% chance that our result is not caused by the subtraction of glitches but is instead caused by random chance. To reduce the error in the computed sensitivity ratio a larger injection set test would be required, this would need a large time and computational power investment which we do not believe is justified in the case of such a minor increase in the sensitivity.

\section{\label{sec:apdx_injections_table}}

Here we have three tables which contain data on the newly-found and newly-missed injections for each injection set. The tables are separated into the values for the inverse false-alarm rate and ranking statistic found for each injection in both searches then the signal-to-noise ratio, $\chi^{2}$ and PSD variation values for each detector in both searches. The first table is the results of the binary black hole injection set, the second table is the binary neutron star injection set and the third is the neutron star black hole injection set. A horizontal dashed line separates newly-found and newly-missed injections, using a false-alarm rate threshold of 2 per year. There were 2 newly-found binary black hole injections which were not found at all by the original search and therefore they do not appear in the first table.

There have been 34 newly-found or newly-missed \gw{} injections when subtracting \scl{} glitches, it is informative to understand how the \gw{} search is influenced by the glitch subtraction to cause this outcome. The ranking statistic~\cite{Davies:2020tsx} represents the legitimacy of a signal being astrophysical in origin and is partially computed using the re-weighted signal-to-noise ratio, which is itself computed using the initial signal-to-noise ratio alongside the various \gw{} discriminators and the PSD variation measurement~\cite{Mozzon_2020}. We use the trigger information saved by the \gw{} search to identify why injections that weren't found previously have been found post-glitch subtraction, and vice versa. 

As an example, we take the smallest false-alarm rate (1 per $8643.73$ years), newly-found, binary black hole injection, and look at the ranking statistic, signal-to-noise ratio, $\chi^{2}$ and, PSD variation measurements in both detectors and both searches -- these values can be found in table~\ref{tab:apdx_changed_snr_bbh}. This injection was originally seen with a false-alarm rate of 100 per year, far above our threshold, a very large increase in the ranking statistic from 13.29 to 27.01 is certainly responsible for the decreased false-alarm rate. There were no changes in the values measured by the LIGO-Livingston detector between searches which is expected as no \scl{} glitches were found within $512$ seconds of the injection. The LIGO-Hanford detector sees a small increase in the signal-to-noise ratio measured, from 7.98 to 8.22, a small increase in the $\chi^{2}$ value, from 2.32 to 2.48, and a very significant decrease in the PSD variation measurement, from 3.47 to 1.50. Using equation 18 of~\cite{Mozzon_2020}, we can calculate a re-weighted signal-to-noise ratio of $4.60$ for the original search and $5.67$ for the glitch-subtracted search, a significant increase in the signal-to-noise ratio. Similar analyses for all the newly-found and newly-missed injections can be made using information found in tables~\ref{tab:apdx_changed_snr_bns} and~\ref{tab:apdx_changed_snr_nsbh}.

The decrease in the PSD variation is true for the three newly-found very low false-alarm rate injections, accompanied by the small changes in signal-to-noise ratio and increase in the $\chi^{2}$ measurement. For newly-found and newly-missed injections which lie close to the 2 per year false-alarm rate threshold there is no definitive reason as to why these injections changed state.

\newpage
\newgeometry{left=2cm,right=2cm,top=2cm,bottom=2cm} 
\begin{landscape}
\begin{table}[tb]
\centering
\caption{\label{tab:apdx_changed_snr_bbh}This table contains the trigger information for the newly-found and newly-missed \textbf{binary black hole injections} recorded by the original search~\cite{gwtc3} and the glitch-subtracted search.} 
\begin{tabular}{|c|c|c|c|c|c|c|c||c|c|c|c|c|c|c|c|}
\hline
\multicolumn{8}{|c||}{Glitch-Subtracted} & \multicolumn{8}{c|}{Original Search} \\
\hline
\multicolumn{2}{|c|}{} & \multicolumn{3}{c|}{H1} & \multicolumn{3}{c||}{L1} & \multicolumn{2}{c|}{} & \multicolumn{3}{c|}{H1} & \multicolumn{3}{c|}{L1}\\
\hline
IFAR & Ranking & SNR & $\chi^{2}$ & PSD & SNR & $\chi^{2}$ & PSD & IFAR & Ranking & SNR & $\chi^{2}$ & PSD & SNR & $\chi^{2}$ & PSD \\ &

Stat. & & & Var. & & & Var. & & Stat. & & & Var. & & & Var.\\
\hline
8643.73 & 27.01 & 8.22 & 2.48 & 1.50 & 8.05 & 2.07 & 1.00 & 0.01 & 13.29 & 7.98 & 2.32 & 3.47 & 8.05 & 2.07 & 1.00 \\
7633.84 & 26.13 & 8.18 & 2.25 & 1.22 & 7.58 & 1.46 & 1.03 & 0.37 & 16.63 & 8.19 & 1.87 & 2.17 & 7.59 & 1.45 & 1.02 \\
4.89 & 19.06 & 7.23 & 1.88 & 0.97 & 10.34 & 3.15 & 1.05 & 1.56 & 18.01 & 7.24 & 2.11 & 0.97 & 10.34 & 3.42 & 1.04 \\
3.14 & 18.65 & 5.89 & 1.54 & 1.01 & 8.86 & 2.40 & 1.09 & 1.52 & 17.99 & 5.89 & 1.54 & 1.01 & 8.86 & 2.40 & 1.09 \\
2.75 & 18.53 & 8.50 & 1.63 & 0.99 & 6.80 & 1.37 & 1.05 & 1.51 & 17.98 & 8.51 & 1.56 & 0.98 & 6.80 & 1.37 & 1.05 \\
2.39 & 18.41 & 7.83 & 2.00 & 1.01 & 8.41 & 3.41 & 1.02 & 1.06 & 17.65 & 7.81 & 2.14 & 1.01 & 8.41 & 3.41 & 1.02 \\
2.18 & 18.32 & 7.02 & 1.90 & 1.02 & 6.35 & 2.15 & 1.01 & 1.79 & 18.14 & 7.05 & 1.92 & 1.02 & 6.35 & 2.16 & 1.01 \\
2.01 & 14.11 & 7.94 & 1.89 & 1.01 & 5.67 & 1.95 & 1.16 & 1.90 & 14.14 & 7.94 & 1.90 & 1.01 & 5.07 & 0.00 & 1.02 \\
\hdashline
1.76 & 18.12 & 6.89 & 1.43 & 1.02 & 6.76 & 2.59 & 0.97 & 3.33 & 18.71 & 6.86 & 1.82 & 1.02 & 6.68 & 2.53 & 0.97 \\
1.79 & 18.14 & 6.40 & 1.69 & 1.02 & 7.56 & 2.01 & 1.01 & 2.05 & 18.27 & 6.40 & 1.69 & 1.02 & 7.56 & 2.00 & 1.01 \\
1.65 & 18.06 & 5.14 & 0.00 & 0.91 & 7.57 & 1.57 & 1.02 & 2.05 & 18.27 & 5.14 & 0.00 & 0.91 & 7.57 & 1.57 & 1.02 \\
\hline
\end{tabular}
\end{table}
\end{landscape}
\restoregeometry 

\newpage

\newgeometry{left=1cm,right=1cm,top=2cm,bottom=2cm} 
\begin{landscape}
\begin{table}[tb]
\centering
\caption{\label{tab:apdx_changed_snr_bns}This table contains the trigger information for the newly-found and newly-missed \textbf{binary neutron star injections} recorded by the original search~\cite{gwtc3} and the glitch-subtracted search.} 
\begin{tabular}{|c|c|c|c|c|c|c|c||c|c|c|c|c|c|c|c|}
\hline
\multicolumn{8}{|c||}{Glitch-Subtracted} & \multicolumn{8}{c|}{Original Search} \\
\hline
\multicolumn{2}{|c|}{} & \multicolumn{3}{c|}{H1} & \multicolumn{3}{c||}{L1} & \multicolumn{2}{c|}{} & \multicolumn{3}{c|}{H1} & \multicolumn{3}{c|}{L1}\\
\hline
IFAR & Ranking & SNR & $\chi^{2}$ & PSD & SNR & $\chi^{2}$ & PSD & IFAR & Ranking & SNR & $\chi^{2}$ & PSD & SNR & $\chi^{2}$ & PSD \\ &
Stat. & & & Var. & & & Var. & & Stat. & & & Var. & & & Var.\\
\hline
5.12 & 19.12 & 5.86 & 1.99 & 1.01 & 7.15 & 1.94 & 1.02 & 1.49 & 17.96 & 5.74 & 2.14 & 1.01 & 7.15 & 1.94 & 1.02 \\
4.91 & 19.07 & 5.48 & 2.29 & 1.01 & 7.92 & 2.16 & 1.01 & 1.37 & 17.89 & 5.78 & 2.25 & 1.01 & 7.23 & 2.13 & 1.01 \\
4.54 & 18.99 & 4.82 & 0.00 & 1.07 & 8.68 & 2.05 & 1.06 & 1.66 & 18.07 & 4.82 & 0.00 & 1.14 & 8.68 & 2.05 & 1.06 \\
2.46 & 18.43 & 5.48 & 1.95 & 1.05 & 7.61 & 2.03 & 0.99 & 1.78 & 18.13 & 5.48 & 1.94 & 1.05 & 7.62 & 2.08 & 0.99 \\
2.34 & 18.39 & 6.25 & 2.18 & 1.01 & 6.82 & 1.96 & 1.00 & 1.45 & 17.94 & 6.09 & 2.05 & 1.01 & 6.83 & 2.01 & 1.00 \\
\hdashline
1.24 & 17.71 & 6.58 & 2.14 & 1.01 & 6.89 & 2.48 & 1.01 & 17.81 & 20.19 & 6.58 & 2.15 & 1.01 & 7.16 & 2.26 & 1.01 \\
1.14 & 17.72 & 6.00 & 2.67 & 0.98 & 8.07 & 2.66 & 1.22 & 7.14 & 19.44 & 6.17 & 2.45 & 0.97 & 8.26 & 2.80 & 1.22 \\
1.89 & 18.19 & 6.32 & 2.19 & 1.00 & 6.79 & 1.83 & 0.99 & 3.57 & 18.78 & 6.37 & 2.12 & 0.99 & 6.79 & 1.75 & 0.99 \\
0.88 & 17.46 & 6.88 & 2.14 & 0.99 & 6.19 & 2.12 & 0.99 & 2.33 & 18.39 & 6.87 & 2.07 & 0.99 & 6.19 & 1.99 & 0.99 \\
1.36 & 17.80 & 6.20 & 2.02 & 1.00 & 6.83 & 2.35 & 1.00 & 2.07 & 18.18 & 6.20 & 2.02 & 1.00 & 6.83 & 2.28 & 1.00 \\
\hline
\end{tabular}
\end{table}
\end{landscape}
\restoregeometry 

\newpage

\newgeometry{left=1cm,right=1cm,top=2cm,bottom=2cm} 
\begin{landscape}
\begin{table}[tb]
\centering
\caption{\label{tab:apdx_changed_snr_nsbh}This table contains the trigger information for the newly-found and newly-missed \textbf{neutron star black hole injections} recorded by the original search~\cite{gwtc3} and the glitch-subtracted search.} 
\begin{tabular}{|c|c|c|c|c|c|c|c||c|c|c|c|c|c|c|c|}
\hline
\multicolumn{8}{|c||}{Glitch-Subtracted} & \multicolumn{8}{c|}{Original Search} \\
\hline
\multicolumn{2}{|c|}{} & \multicolumn{3}{c|}{H1} & \multicolumn{3}{c||}{L1} & \multicolumn{2}{c|}{} & \multicolumn{3}{c|}{H1} & \multicolumn{3}{c|}{L1}\\
\hline
IFAR & Ranking & SNR & $\chi^{2}$ & PSD & SNR & $\chi^{2}$ & PSD & IFAR & Ranking & SNR & $\chi^{2}$ & PSD & SNR & $\chi^{2}$ & PSD \\ &
Stat. & & & Var. & & & Var. & & Stat. & & & Var. & & & Var.\\
\hline
9961.55 & 27.41 & 5.55 & 2.08 & 1.23 & 10.26 & 2.27 & 1.12 & 0.01 & 12.83 & 6.01 & 2.39 & 2.25 & 8.54 & 2.37 & 1.12 \\
141.35 & 22.28 & 7.76 & 2.20 & 1.02 & 6.54 & 2.35 & 1.01 & 1.21 & 17.77 & 7.83 & 1.94 & 1.02 & 7.29 & 2.50 & 1.00 \\
18.26 & 20.36 & 5.94 & 2.20 & 1.09 & 6.02 & 1.99 & 1.10 & 1.29 & 17.82 & 5.74 & 2.05 & 1.09 & 6.27 & 2.10 & 1.18 \\
17.12 & 20.29 & 7.64 & 2.00 & 1.11 & 6.66 & 2.15 & 1.05 & 1.37 & 17.89 & 7.36 & 2.35 & 1.11 & 6.66 & 2.15 & 1.05 \\
12.94 & 20.03 & 6.52 & 1.79 & 1.07 & 7.31 & 2.32 & 1.13 & 1.96 & 18.23 & 6.57 & 1.85 & 1.20 & 7.31 & 2.32 & 1.13 \\
7.12 & 19.42 & 5.13 & 0.00 & 1.05 & 8.33 & 2.08 & 1.04 & 0.03 & 14.23 & 5.30 & 2.16 & 1.05 & 7.44 & 2.43 & 1.04 \\
3.61 & 18.77 & 5.80 & 2.01 & 1.02 & 7.43 & 1.88 & 0.93 & 1.81 & 18.15 & 5.80 & 2.01 & 1.02 & 7.43 & 1.97 & 0.93 \\
2.15 & 18.22 & 7.41 & 2.04 & 1.02 & 5.57 & 2.03 & 1.03 & 0.00 & 4.34 & 5.64 & 2.18 & 0.98 & 5.04 & -0.00 & 1.01 \\
\hdashline
0.23 & 16.16 & 5.79 & 1.81 & 0.96 & 8.78 & 3.93 & 1.03 & 11.69 & 19.93 & 5.78 & 1.89 & 0.96 & 8.79 & 3.17 & 1.03 \\
1.56 & 17.92 & 6.71 & 1.91 & 0.99 & 6.73 & 2.50 & 1.05 & 2.68 & 18.41 & 6.71 & 1.90 & 0.99 & 6.72 & 2.39 & 1.05 \\
1.72 & 18.10 & 5.77 & 1.90 & 0.99 & 6.77 & 1.74 & 1.00 & 2.08 & 18.28 & 5.77 & 1.90 & 0.99 & 6.80 & 1.69 & 1.00 \\
\hline
\end{tabular}
\end{table}
\end{landscape}
\restoregeometry 

\end{document}